\documentclass[acmlarge,screen,table,xcdraw,authorversion]{acmart}
\newcommand{\imagepath}[1]{images/cropped/#1}
\newcommand{\figpath}[1]{figures/#1}

\usepackage{multirow}

\newcommand{\imagen}{Envisage} 
\newcommand{\Imagen}{\imagen{}}
\newcommand{\IS}{Image Search}

\newcommand{\prompt}[1]{\textquotedbl\texttt{#1}\textquotedbl}

\usepackage[T1]{fontenc}
\usepackage{lscape}

\usepackage{xparse}
\usepackage{enumitem}
\setlist[description]{
  font={\sffamily\bfseries},
  labelsep=5pt,
  labelindent=20pt,
  labelwidth=\transcriptlen,
  leftmargin=\transcriptlen,
}

\newlength{\transcriptlen}

\NewDocumentCommand {\setspeaker} { mo } {%
  \IfNoValueTF{#2}
  {\expandafter\newcommand\csname#1\endcsname{\item[#1:]}}%
  {\expandafter\newcommand\csname#1\endcsname{\item[#2:]}}%
  \IfNoValueTF{#2}
  {\settowidth{\transcriptlen}{#1}}%
  {\settowidth{\transcriptlen}{#2}}%
}

\usepackage{caption}
\usepackage{subcaption}

\AtBeginDocument{%
  \providecommand\BibTeX{{%
    Bib\TeX}}}

\setcopyright{acmcopyright}
\copyrightyear{2022}
\acmYear{2022}
\acmDOI{XXXXXXX.XXXXXXX}

\acmConference[Conference acronym 'XX]{Make sure to enter the correct
  conference title from your rights confirmation email}{June 03--05,
  2018}{Woodstock, NY}
\acmPrice{15.00}
\acmISBN{978-1-4503-XXXX-X/18/06}

\usepackage{soul}

\begin{document}

\title[Prompts as AI Design Material]{A Word is Worth a Thousand Pictures: Prompts as AI Design Material}


\author{Chinmay Kulkarni}
\affiliation{%
  \institution{Google, Inc.}
  \city{Atlanta, Georgia}
  \country{United States}}
\email{ckulkarni@google.com}

\author{Stefania Druga}
\affiliation{%
  \institution{Google, Inc.}
  \city{Mountain View, California}
  \country{United States}}
\email{druga@google.com}

\author{Minsuk Chang}
\affiliation{%
  \institution{Google, Inc.}
  \city{Seattle, Washington}
  \country{United States}}
\email{misukchang@google.com}

\author{Alex Fiannaca}
\affiliation{%
  \institution{Google, Inc.}
  \city{Seattle, Washington}
  \country{United States}}
\email{afiannaca@google.com}

\author{Carrie Cai}
\affiliation{%
  \institution{Google, Inc.}
  \city{Mountain View, California}
  \country{United States}}
\email{druga@google.com}

\author{Michael Terry}
\affiliation{%
  \institution{Google, Inc.}
  \city{Seattle, Washington}
  \country{United States}}
\email{michaelterry@google.com}

\renewcommand{\shortauthors}{Kulkarni, et al.}
\newcommand{\LPIM}{text-to-image model}
\newcommand{\LPIMs}{text-to-image models}
\begin{abstract}
Recent advances in Machine-Learning have led to the development of models that generate images based on a text description. Such large prompt-based text to image models (TTIs), trained on a considerable amount of data, allow the creation of high-quality images by users with no graphics or design training. This paper examines the role such TTI models can play in collaborative, goal-oriented design. Through a within-subjects study with 14 non-professional designers, we find that such models can help participants explore a design space rapidly and allow for fluid collaboration. We also find that text inputs to such models (“prompts”) act as reflective design material, facilitating exploration, iteration, and reflection in pair design. This work contributes to the future of collaborative design supported by generative AI by providing an account of how \LPIMs{} influence the design process and the social dynamics around design and suggesting implications for tool design.

\end{abstract}





\begin{teaserfigure}
\includegraphics[width=\textwidth]{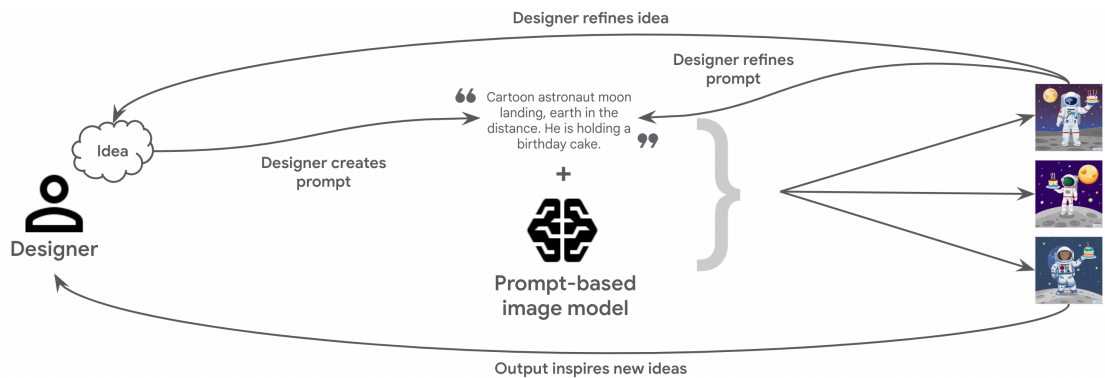}
\caption{Large prompt-based text-to-image models enable rapid exploration of a design space, and fluid collaboration. By allowing users to declaratively and quickly create images through text descriptions, these text prompts act as a \textit{reflective design material} aiding exploration and collaboration. We observe that designers create prompts based on their tacit understanding of the model, and model outputs in turn both spark new ideas and allow rapid refinement of prompts.}
\label{teaserfigure}
\end{teaserfigure}

\maketitle

\section{Introduction} 
Recent advances in  \LPIMs{} (TTI models), allow users to generate high-quality images based on a text description or ``prompt''. In ways similar to large language models (LLMs), where increasing size of models has discontinuous benefits~\cite{wei2022emergent}, the increased size of recent \LPIMs{} has yielded discontinuous and qualitative differences in the quality of images produced~\cite{saharia2022Imagen}. The quality of \LPIMs{} has led to energetic communities of practice, where enthusiasts readily share designs and prompts. In some cases, results obtained from these models are so good that one artist has even won a prestigious art competition \cite{AIwinsst11:online}. 

The image quality obtained by this new generation of \LPIMs{} enables the research community to rethink significant aspects of the design process in light of these models. For example, questions arise about whether designers could delegate parts of the creative process to a model; how the nature and the style of creative design change with these models; and how designers should best share and collaborate on multimodal (i.e., text and image) creative processes with these models available as design resources. Finally, potential new design processes among non-professional designers are particularly interesting, as TTI models allow those without professional training to create high-quality images easily.


This paper investigates how the creative design processes of non-professional designers change while using TTI models. Since design practices often emerge through social interactions (e.g., designers working together in a studio or sharing their work online), we also investigate how \textit{social interactions} are affected with the introduction of a TTI model. Recent work suggests that generative AI may play a role in influencing social dynamics between pairs of people~\cite{suh2021GenerativeMusic}. With the advent of ``prompting'' as a new form of interacting with AI, we build on this emerging body of literature by investigating how prompting may affect social dynamics during collaborative design.

Our paper examines the following two research questions: 
\begin{itemize}
    \item RQ1: How does using prompt-based image generation change the design process of non-professional designers, especially compared to current tools for finding appropriate images such as web search?
    \item RQ2: How do prompt-based image generation models change collaborative dynamics during design?
\end{itemize}

Note that while prior work on human-AI co-creative systems has uncovered challenges that frequently arise when users co-create with AI (e.g., users must deal with uncertainty in the capabilities of AI capabilities while simultaneously making sense of complex outputs from the system \cite{yang2020DesignAI}), it is unknown how these challenges manifest when users interact with prompt-based i.e., TTI models. It is also unclear what other challenges are unique to working with these models and how we might best address them when designing new interfaces for these systems.

To address these research questions, we conducted a design study with participants from a large technology company who use prompt-based image models in a non-professional capacity. In our controlled study, participants worked in pairs and created graphic designs with or without the assistance of a prompt-based image generation model. We present results in this paper from our direct observations of the designer pairs, conversations during this design session, and post-study interviews. We also compare the artifacts produced by participants for creativity, completeness, and appropriateness to the design brief. 

Overall, we found that TTI models change the design process by allowing designers to create images \textit{declaratively}, i.e., through a simple description of the desired image. This declarative design changes existing design practices in two ways. First, because prompts can capture high-level image descriptions, TTI models allow faster exploration of the design space, potentially leading to more creative design. {(At the same time, we acknowledge that not all creative ideas for generating images can be expressed easily, or at all, in words. As new text-guided image-to-image models become more capable, we expect that they will aid exploration even further.)} Second, because prompts are text, prompt-based image generation leads to easier sharing of design ideas, allowing designers to collaborate and build on each others' work more successfully.  

Throughout our study, we observed that prompts played a central role in the design process and collaboration. This leads us to argue that prompts act as a \textit{reflective design material} in the design process. Specifically, we found that our participants developed a \textit{tacit}, rather than technical, understanding of how different aspects of prompts (such as specific keywords) influence the image generated. 

In addition, prompts enabled fluid collaborations between participants as they shared, modified, and iteratively improved each others' prompts. The ability to easily edit and refine an image via a text interface made the design process more fluid, and sharing prompts easily aided collaboration, which uniquely placed prompts as design material in a multimodal creative setting.

Prompts also allowed participants to engage in reflective practice with the AI model. Specifically, the \LPIM{} outputs made it seem like a somewhat opinionated ``design partner'' in its preference to generate certain kinds of images, or when the images generated were wildly different from the prompt's intent or were challenging to modify. Participants leveraged prompts to reflect on the model's results  
and engaged in reflective conversations with each other through their prompts, envisioning novel designs they did not foresee~\cite{schon1984architectural,ghajargar2018thinking}. This leads us to suggest that prompts act as reflective design materials (``objects to think with'' \cite{turkle2005second}).

In sum, this paper makes the following contributions:
\begin{itemize}
 
\item {\textbf{Changes in the design process.} We articulate ways in which TTI models change the design process. Specifically, models changed design processes through faster iteration, creating images of novel ideas and unlikely combinations (e.g., a giraffe in a Lamborghini) quickly. They also changed the design process because they rely on \textit{indirect} manipulation of images (through text). Together, this led to wide-ranging exploration where the models rapidly ``filled in'' many unspecified details of the images.  However, participants also struggled to control low-level factors (e.g., cropping, position, text) in the generated image. In sum, users perceived their work to be significantly more creative when using a TTI model compared to image search. However, external raters found no significant difference. }

\item {\textbf{Changes in collaborative practices.} We identify ways prompts changed collaborative dynamics during design. While prompts empowered collaborators to combine multiple people's disparate ideas fluidly, their non-determinism hindered coordination. In addition, because participants frequently shared images without the prompts that generated them, partners had asymmetric access to TTI models.}

\item \textbf{Prompts as a design material.} {Given these empirical results, we  conceptualize prompts as a new design material that both enables rapid exploration of a design space and modulates collaboration, but one that is imperfect, especially for fine-grained control.} 
\end{itemize}

\section{Related Work}
This paper builds on three areas of related work: large machine-learning models generally, Human-AI collaboration issues, and the use of AI to support creativity and design. We briefly outline significant findings from prior work below and how this paper extends these findings.

\subsection{Large machine learning models and their use in rapid prototyping}
Recent advances in deep neural networks have enabled the creation of large language models (LLMs) capable of generating highly realistic language output \cite{NEURIPS2020_1457c0d6, lieberjurassic}. When provided with brief contextual information, such as a textual description of the task, these models can mimic the performance of models specially trained for particular tasks such as classification (``Classify whether this review is positive or negative"), question answering (``Given the information below, answer the following questions") and summarization (``Summarize this article''),  i.e., they act as zero- or few-shot learners for a wide array of problems \cite{betz2021Aloud, liu2021InContext, lu2021PromptOrder}. Designing the form of contextual information to provide to the model is referred to as \textit{prompt programming} or \textit{prompt engineering} \cite{wu2022AIChains}. Our work adds to this rich scholarship by examining how designers can engineer prompts to generate images collaboratively. While prior work in this space has generally focused on identifying new capabilities of prompt engineering, our study focuses primarily on how these new capabilities are used.

A parallel line of research has explored generative models for images \cite{gan2014Goodfellow, Zhu_2017_ICCV, Karras_2019_CVPR, gan2018Brock}. Critically, Mansimov et al. \cite{mansimov2015ImgFromCaption} showed that generative image models conditioned on image captions can generate images from natural language input. Leveraging the technological advances in LLMs, a significant number of large prompt-based image-generation models have since been created in the last two years -- including DALL·E \cite{pmlr-v139-ramesh21a, dalle1Web} and DALL·E 2 \cite{ramesh2022DALLE2, dalle2Web}, Parti \cite{yu2022Parti, partiWeb} and Imagen \cite{saharia2022Imagen, imagenWeb}, Stable Diffusion \cite{Rombach_2022_CVPR, stableDiffusionWeb}, and Midjourney \cite{midjourneyWeb}. Given the ability of these \LPIMs{} to generate highly detailed images in a vast array of styles based on the user's prompt, these models present a significant opportunity for application to creative visual tasks \cite{bommasani2021Foundation}. In this work, we investigate the use of text-to-image model's impact on collaborative design tasks.

\subsection{Human-AI co-creation}
The question of what role AI systems play in mediating communication between collaborators \cite{hancock2020AICommunication}, and how AI systems can act as teammates has been widely debated \cite{seeber2020AITeammates,wang2020HumanAi}. Prior work has found that while Human-AI collaboration can improve the abilities of unassisted humans and AI systems without human input, designing such collaborative systems remains challenging~\cite{yang2020profiling}. Particular challenges are a lack of clarity on the capabilities and limitations of AI,  the complexity of output of AI systems, and the role of randomness in AI outputs, both of which impede a designerly understanding of AI~\cite{yang2020DesignAI}. To this literature, we contribute an account of how \textit{prompts} emerge as a reflective design material that help non-professional designers explore the design space of possibilities, and collaborate with each other.

Designing \textit{co-creative} systems (as opposed to decision support systems, for instance) have particular challenges~\cite{buschek2021nine}. Key to our current study is the potential for AI systems to be a ``time-waster'' that user interfaces may impose a “bottleneck” on creative use of the AI, and that AI provides overwhelming amount/detail of content that distracts or creates choice-overload~\cite{buschek2021nine}, resulting in poor design.

This paper enriches this literature by describing how pairs of non-professional designers navigate a \LPIM{}. In this respect, they extend our understanding of interacting with \textit{multimodal} AI co-creation. At the same time, multimodal systems have long been known to improve users' expressive power and efficiency~\cite{oviatt2000multimodalnatural}, our findings on how AI models such as \LPIMs{} can modulate expressivity and efficiency.

\subsection{Collaborative Creativity Support and the use of AI}
Tools that support creative tasks have been widely studied in the fields of HCI and psychology \cite{frih2019CSTinHCI}, both for individuals and groups, and from both a human-centric perspective and a computational perspective \cite{kantosalo2016modes}. Given the focus of this paper on collaborative design, we focus this section on creativity support tools for groups (see~\cite{wang2017literature} for an overview of tools for individual support). Tools supporting groups collaborating on creative tasks have studied both the creative process and the communities of practice in which such processes are situated~\cite{ScratchI75:online, Dynamicl51:online, bruckman1998community}. For example, Scratch~\cite{ScratchI75:online} supports collaborative game development through a block-based programming language and supports self-directed discovery-based learning among children~\cite{roque2016supporting}. The critical learning mechanism in Scratch beyond self-directed discovery is ``re-mixing'' or a creative bricolage of other creators' ideas~\cite{dasgupta2016remixing}. Similarly, online communities, such as Dribbble, allow designers to find and build on collaborators' ideas as a critical mechanism to improve creative output~\cite{bruckman1998community,marlow2014rookie}. This paper extends this scholarship by describing how collaborating designers re-mix each others' ideas when creating AI-assisted images.

In recent years, creativity support tools that use AI have been growing in popularity and have been studied in domains such as creative writing \cite{clark2018Writing, gero2019Metaphoria}, music creation \cite{mccormack2019Music, louie2020Music, huang2020AISong, suh2021GenerativeMusic}, and drawing \cite{davis2015Drawing, davis2016Drawing, oh2018Drawing, karimi2019Drawing}, and design ideation \cite{koch2019Ideation, Sbai_2018_ECCV_Workshops, jeon2021Fashion, quanz2020machine}. In addition, several systems have been designed to support the collaborative generation of creative content. Together, studies of these systems suggest that AI models can shape interactions between collaborating partners and individual and group cognitive processes. This work informs our focus on how \LPIMs{} change collaboration during design and extends the scholarship on how multimodal AI models (i.e., using text to generate images) influence collaboration.

\section{\LPIM{}}
Participants in our study used a \LPIM{} that has not been released to the public. This model, \Imagen{}, is a text-to-image diffusion model that achieves a very high degree of photorealism and a deep level of language understanding (comparable in quality to DALL-E 2 \cite{ramesh2022DALLE2,dalle2Web}, Imagen \cite{imagenWeb}, and Parti \cite{partiWeb}). {The model is a "diffusion model". Diffusion models are trained by first destroying training data through the successive addition of noise, and then learning to recover the data by reversing this noising process. After training, a diffusion model can be used to generate data by simply passing sampled noise through the learned de-noising process. To guide the reconstruction trajectory, more recent implementations of diffusion models use text, semantic maps, or other images to condition what possible image should be generated (reconstructed) from the space of all possible options, with different probabilities (also called the latent space). Our model uses text to guide reconstruction.} In particular, it uses a generic large language model, trained on text-only corpora, for encoding text inputs (prompts), and a specially-trained diffusion model for generating images. 

{While the objective of this paper is not to study a specific TTI model, to help contextualize our results, we note that previous work demonstrated that the \Imagen{} model exhibits state-of-the-art performance on automated metrics (FID), human-rater comparable performance on text-image alignment to the MS-COCO dataset itself, and a high degree of human-perceived photorealism in generated images (redacted reference).}
%
While \Imagen{} can often produce legible, correctly spelled text in images, it is not guaranteed to do so. In addition, generated images are constrained to be square and of fixed resolution and size.

Participants used \Imagen{} through a web-based user-interface. This web-based UI has one text box for the input text (prompt) and generates eight candidate images at a time (presented in two rows as a $4x2$ grid.) Generating these images takes approximately 20 seconds once a prompt is entered. The UI currently does not record a history of past prompts used. Because \Imagen{} uses a random seed as input (which is not visible to the user), users see different image results on consecutive runs of the model, even when providing an identical prompt input. Both the prompt and the generated images are saved, and a unique URL is generated for every image-generation run, allowing easier sharing. (Visiting this URL reloads the previously generated images.) The current UI does not have any other collaborative features. {Finally, the authors of this paper had no involvement in the creation of the model or the interface used in the study.}

Our study focused on the practices of non-professional designers. We are  motivated to study how practices of non-professional designers might change particularly because TTIs may empower those without professional artistic or design training to create images with a textual description alone rapidly.
TTIs may also allow non-professional designers  to rapidly prototype, see and borrow from examples, and share multiple designs, accruing benefits that professional designers obtain from these practices~\cite{dow2010parallel}.

\subsection{Participants}
Participants in our study were invited from a pool of employees at a large, US-based tech corporation who filled out a survey suggesting that they have used \Imagen{} or similar models in the past. 

It must be noted here that our study makes a distinction between \emph{design practices} -- i.e. the individual and collaborative cognitive processes common to the task of design, and the practices of professional designers, which are learned through participation in this community of practice. Our paper is focused on the former as we are interested in the effects of TTI models among non-professional designers. 

As a result, {we filtered this survey to find only those who had been exposed to these models outside their job responsibilities and who were not professional designers (such as UX designers, visual designers, etc.). Specifically, to target non-professionals, we narrowed our pool to participants who reported that they had interacted with these models as ``part of your creative work pipeline'' or ``out of curiosity (not work-related)''. For practical reasons, recruiting those who had some initial exposure to the model also allowed us to observe the task within the scope of the 1-hour study. }

From this pool, we invited 16 participants across the company to participate in our design study. 14 participated (11 identified as male, three as female.) {Participants had a variety of job roles, from sales/marketing, software development, and project management. We sought to balance the participant pool to include both tech and non-tech participants.} Two participants did not fill out the pre-study or post-study survey but fully participated in our study.  We present the results of the survey without these participants. Participants were given a company-internal gift card {valued at US \$30)} for participating in the study.

Two expert visual designers, both identifying as female (working as UX designers in the company), rated participants' final outputs blind to experimental conditions and participant identity.

{Because the model was not released to the public (at the time the study was run), we were limited to employees working at our organization. Despite this limitation, the deployment allowed us to gain fruitful insights from people across a range of different job roles, including non-tech roles. As TTI models become more widely available, future work should examine the generalizability of our results.}

\subsection{Study design}

\subsubsection{\IS{} as a baseline}
{To find a reasonable baseline condition to compare against how participants used TTIs, we conducted pilot experiments to find existing design practices among non-professional designers.  We found that participants overwhelmingly started with searching for images online, and if necessary, editing these images in slide decks or similar software.} 

While reasons varied (from thinking of returned images as inspiration or as building blocks to be used right away), \IS{} {enables users to find images that depicted what they wanted using high-level, natural language descriptions. Like TTI, Image Search also uses text as input.


While a generative image-editing tool might seem like a more natural comparison to generative TTI models than} \IS{}{, which only surfaces pre-existing images, notably none of our pilot participants used generative image-editing tools (e.g. Adobe Illustrator) or machine learning-powered image generation (e.g. GANs). Given the complexity of existing image-generation tools like Illustrator relative to TTI, such a comparison might also make this comparison unfair.
}
\subsubsection{Procedure}
Our study was conducted entirely online, with participants using Google Meet. Participants shared their videos so they could see each other and their active window to see each others' work. In addition to the participant, one or two experimenters joined the video call. Participants were asked to share their browser window to allow both the experimenter and their partner to follow their progress. Before the start of the experiment, participants filled out a short demographic survey after consenting to participation. {Before proceeding with the study, we reminded participants that we had no involvement in the creation of the model or the interface used in the study, and that we welcomed their honest reflections throughout the study.}

Participants were assigned to a design partner at the start of the study. Our study was designed as a within-subjects study with two design sessions. In one design session, in the \IS{} condition, participant pairs were allowed only to use \IS{} (with Google \IS{}) -- participants could perform an unlimited number of queries, and use resulting images in their invitation. In the other design session, in the \Imagen{} condition, they were additionally allowed to use \Imagen{}; participants could similarly generate images for an unlimited number of prompts, and use resulting images. Each design session lasted 20 minutes, in which participants were asked to create a complete design. The two experimental conditions were counterbalanced, so half the pairs of participants completed designs with \IS{} alone first, and the other half completed designs with additional access to \Imagen{}.

Throughout the study, participants worked with the same assigned design partner. In the first design session, the pair of participants collaboratively designed a party invitation to the birthday party of Alice from \textit{Alice in Wonderland}. In the second session, the pair of participants collaboratively designed a party invitation to celebrate the 55th anniversary of the first Moon Landing. Appendix~f{app:task-descroption} includes both design briefs in full. Because \Imagen{} does not allow the creation of images that include photo-realistic people, participants were told not to use photo-realistic images of people in their design. Further, to respect creator rights and to simulate a realistic design task, participants were also not allowed to use images that were under copyright. (Images that were public domain or licensed under a Creative Commons license were allowed -- the \IS{} interface has filters to search for such images.)~{This restriction is similar to prior work in this area (e.g.}~\cite{dow2010parallel}{.)}

In both sessions, participants created their invitations in Google Slides. Using Slides, participants could for instance crop images, compose multiple images into a single composition, and add text. Google Slides does not currently have features to remove backgrounds from images, or perform other image editing, such as adding filters. We chose to use Google Slides rather than a professional tool such as Adobe Illustrator because we wanted to study the practices of non-professional designers. Furthermore, Google Slides is used extensively in the organization we studied, ensuring that all participants would be familiar with how to use it. Participants were told that their designs would be rated for creativity, appropriateness to the design brief, and completeness and that the best designs would receive a prize.

After the study, each participant privately filled out a survey self-assessing their design along the dimensions of creativity, appropriateness, and completeness (based on survey scales adapted from~\cite{dow2010parallel}) {(Modifications: tailor language to invitations instead of advertisements in the original study, replace adherence to the ``client's theme'' with adherence to the theme, omit questions related to graphic design principles such as typographic balance because that was not the focus of our study)}. We also asked participants about their collaborative experience (based on ``relationship conflict'' scales from \cite{jehn2001dynamic}). We also asked two expert designers (who did not participate in the study and were not involved in the research) to rate participants' designs along these dimensions. These raters used the same scales as participants and were blind to the condition. {(Each rater rated all designs produced.)}

\subsection{Data collection and analysis}
Our study resulted in pre-study survey data and video recordings of all the design sessions. {Recordings were automatically transcribed, and corrected by either of the first two authors.} The {first two} authors analyzed the video transcriptions and noted comments on participants' non-verbal interactions for the qualitative analyses. The final corpus included 168 pages of transcripts (48169 words). The first two authors each reviewed the transcript data independently, looking for ways of explaining the experimental conditions. In this process, the authors analyzed each transcript using emic codes that emerged from the study sessions \cite{miles1984drawing,patton1990qualitative}. After a final coding frame was developed, the second author coded all the transcripts. If new codes emerged, the first two authors discussed discrepancies in the analyses until they reached an agreement. The final list of codes, their definitions, and examples is included in the appendix.
This process was used to develop categories, which were then conceptualized into broad themes after further discussion. The two first authors extracted salient themes from the study transcripts and independently generated hypotheses and points of discussions \cite{braun2006using}. Using these data, the authors participated in two interpretation sessions to arrive at the primary themes reported in this paper. {While we are limited by our sample size, we did not observe large differences in participant processes by gender.}

\section{Results}
 In this section, we first present the participants' design outputs, as well as quantitative results and interaction patterns. Then, in the following sections, we describe how \LPIMs{} changed the design process, and affected collaborative dynamics during design.
 
Note that even though participants in the \imagen{} condition were allowed to use images found through  using \IS{}, only one pair of participants did so. To simplify our description of results, we therefore describe them as results while using \imagen{} and while using \IS{}, even though participants using \imagen{} always had access to \IS{}.

\subsection{Design Outputs and Quantitative Findings}
{Qualitative results regarding the processes followed by participants is possibly the larger contribution of our work. However, for the sake of completeness and to offer a statistical overview of participant behaviors and preferences, we briefly mention quantitative results below.}

Nearly all participants spent all the available time during their first session (regardless of whether they used \IS{} or \imagen{}), $M=18.5$ minutes. Many participants used less time in the second design session, regardless of condition, $M=14.2$ minutes, suggesting there was a learning effect in the task. However, there was no significant difference between \IS{} and \imagen{} conditions.   

When working with their partner, participants followed different collaborative styles, which we briefly describe in Section~f{sub:collaboration}. Participants typically brainstormed about the general theme of their design (characters from Alice in Wonderland), and then started to look for images (using \IS), or generated images (using \imagen{}) that fit this theme. While some pairs had one screen shared with one of the participants ``driving'' the integration of images into the final design, while the other looked for images, many pairs  worked collaboratively on editing the prompts to generate images. Finally, some pairs worked in parallel, sharing interesting results with each other. These pairs then collaboratively edited the slide deck to create the final design.
 
Figure~f{fig:cake-prompts} shows some of the final invitations the participants created. Participants' invitations created in the \IS{} condition had an average of $4$ images in the design ($median=4$), while those in the \imagen{} condition had $2.4$ images on average ($median=2$). We discuss possible reasons for this difference in Section~f{ssub:opinionated}.

Participants self-reported their final design to be more creative when they used \imagen{} (5-point Likert scale, mean=3.6) than when they used \IS{} alone (M=3). This improvement (M=0.6) was statistically significant ({two-sided paired} $t(13)=2.65, p=0.02$.) There was no difference in how complete participants reported their creations to be (\IS{} $M=3.6$, \Imagen{} $M=3.2$), or how appropriate it was to the design brief  (\IS $M=4.2$, \Imagen{} $M=4.1$). On the other hand, external raters did not find any significant differences in the creativity, completeness, or appropriateness of the brief for designs in either condition. 

In a post-study survey, participants did not note any significant differences in their ability  ``to manage any relationship tension'' in their work group (paired t-test, $p=0.19$), ``to politely include my partner’s ideas in the final design while also preserving my own'' ({two-sided} paired t-test, $p=0.27$), or ``to decide about who should do what in our group, even when we had some differences in opinion'' ({two-sided} paired t-test, $p=0.16$). However, participants did reveal a preference for using \imagen{} or a similar model were they ``to complete a similar task in the future'' (mean rating $=3.8$, median $=4$, on a 1-5 Likert scale, \textit{5=``Strongly prefer to use \Imagen{}/similar
model''}.)  

Given these self-reported preferences for using \imagen{} and similar models, along with modest differences in the output quality, we focus the rest of this section on the qualitative differences in the processes that participants employed.

\subsection{Interaction Patterns}

During the study, we observed differences in how participants queried or prompted \IS{} and \imagen{}. Participants understood that \IS{} found pre-existing images on the Internet and so used broad queries that they hoped would yield useful results (e.g. \prompt{tea party}.) If these queries did not yield relevant results, participants searched for related terms instead (e.g., \prompt{mad hatter} $\rightarrow$  \prompt{hare with a hat} $\rightarrow$ \prompt{crazy top hats}). As such, participants (correctly) used \IS{} as a \textit{querying} interface. In contrast, participants' inputs to \Imagen{} could best be described not as \textit{queries} but as \textit{descriptions}, such as \prompt{Colorful drawing of a Cheshire cat from Alice in Wonderland. The cat is wearing a birthday hat and is on a white background.} Throughout the rest of this paper, we call these input descriptions \textit{prompts} to distinguish them from queries.

Participants wrote increasingly elaborate prompts with \Imagen{} during their design session, especially when the image results were disappointing. For instance, P7{} and P8{} tried the prompt \prompt{Beach party on the moon, on the moon in the Sea of Tranquility. Digital art.} Unfortunately, \Imagen{} did not generate any images for this query, and participants hypothesized this was because the beach party had nudity or other content that led  \Imagen{} to block it. (In actuality, it is likely these images were blocked because \imagen{} does not generate images with photo-realistic people in them.) These participants then modified their prompt several times, ending with \prompt{A doodle of a beach party of fully suited astronauts on the Moon in the Sea of Tranquility. The Sea of Tranquility has water in it, and some astronauts are surfing in it with surfboards that have the "NASA" logo on them. Digital art.}


%
\begin{figure}
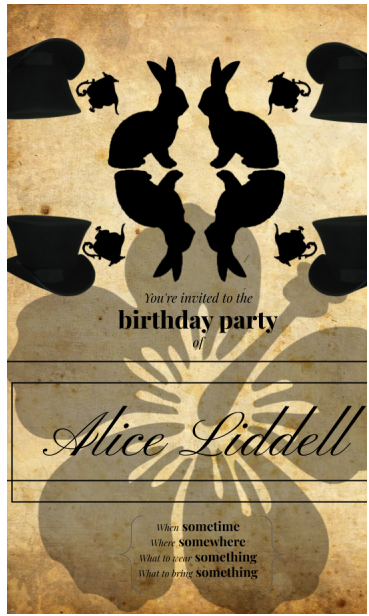
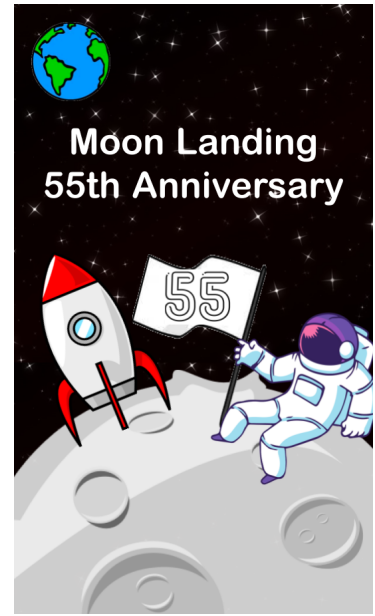
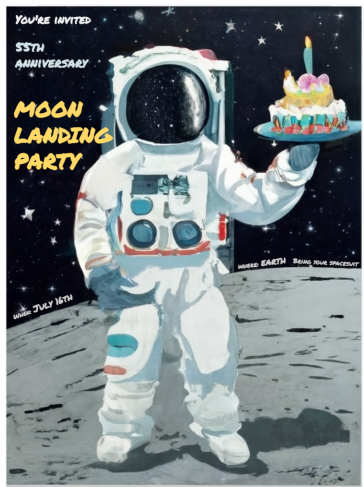
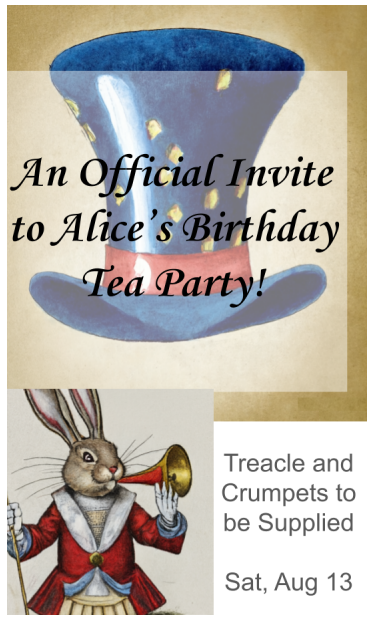
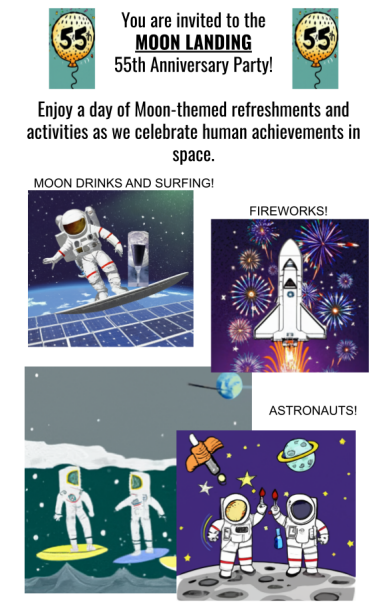

     \centering
     \begin{subfigure}[t]{0.3\textwidth}
         \centering
         \includegraphics[width=\textwidth]{\figpath{image-search/Dave-Ian}}
         \label{fig:image-search-1}
     \end{subfigure}
     \hfill
     \begin{subfigure}[t]{0.3\textwidth}
         \centering
         \includegraphics[width=\textwidth]{\figpath{image-search/Emily-Hendrik}}
         \label{fig:image-search-2}
     \end{subfigure}
     \hfill
     \begin{subfigure}[t]{0.3\textwidth}
         \centering
         \includegraphics[width=\textwidth]{\figpath{image-search/Bardia-Dillon}}
     \end{subfigure}
     
     \begin{subfigure}[t]{0.3\textwidth}
         \centering
         \includegraphics[width=\textwidth]{\figpath{imagen/Emily-Hendrik}}
         \label{fig:imagen-1}
     \end{subfigure}
     \hfill
     \begin{subfigure}[t]{0.3\textwidth}
         \centering
         \includegraphics[width=\textwidth]{\figpath{imagen/dave-ian}}
         \label{fig:imagen-2}
     \end{subfigure}
     \hfill
     \begin{subfigure}[t]{0.3\textwidth}
         \centering
         \includegraphics[width=\textwidth]{\figpath{imagen/peter-justin}}
     \end{subfigure}
        \caption{A few of the designs participants created in our study. \textit{Top row}: Designs without access to prompt-based image model; \textit{Bottom row}: Designs with access to prompt-based image model. {(In both conditions, designs shown are ones with the highest overall rating by independent experts.)}}
        \label{fig:cake-prompts}
\end{figure}

\begin{figure}
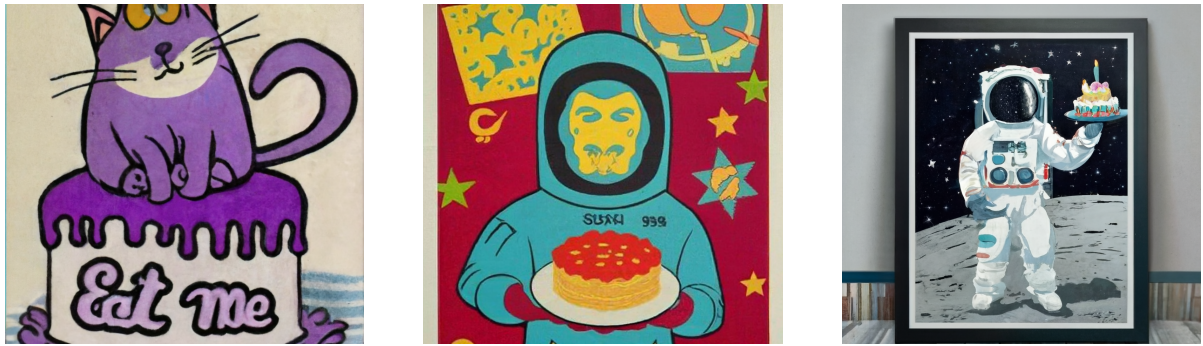

     \centering
     \begin{subfigure}[t]{0.3\textwidth}
         \centering
         \includegraphics[width=\textwidth]{\imagepath{cake3.png}}
         \label{fig:cake1}
     \end{subfigure}
     \hfill
     \begin{subfigure}[t]{0.3\textwidth}
         \centering
         \includegraphics[width=\textwidth]{\imagepath{cake2.png}}
         \label{fig:cake2}
     \end{subfigure}
     \hfill
     \begin{subfigure}[t]{0.3\textwidth}
         \centering
         \includegraphics[width=\textwidth]{\imagepath{cake1.png}}
         \label{fig:cake3}
     \end{subfigure}
        \caption{A few of the images participants created with \imagen{}. As can be seen, \imagen{} does not always generate images that are properly cropped; participants used prompts such as \prompt{...framed art} to generate images with better composition (far right).}
        \label{fig:cake-prompts}
\end{figure}

\subsection{How TTI models changed the design process}\label{sub:collaboration}
Through rapid  image creation and their indirect nature, where images were created through text descriptions, \LPIMs{} led to new design practices, as outlined below. 

\subsubsection{Indirect and rapid image creation through text allowed {new} creative freedom}
Participants noted how creating with \imagen{} was indirect, as it involves ``creating prompts that create images'' (P13{}). This indirectness and the flexibility of prompt editing allowed participants to rapidly explore the design space of alternatives. This was most apparent when participants used \imagen{} to take on other 'artist' personalities, which would otherwise have taken years of practice. P4{} noted:  ``If you made a poster, it [the poster] would have had your style associated with it by default because you have to learn [and develop a particular style]… It is harder to switch between styles. Whereas \Imagen{}, you could just be like `1960s poster’ or like in the style of whoever: Picasso''. At the same time, participants felt faster image generation would allow for even more exploration. P3{} noted: “Because it takes so long to generate a bunch of different images, I didn't really move off to, you know, how else that card could look.”

Participants also noted how the model implicitly steered such rapid exploration. Despite this steering, participants noted how they still remained in control over the ``personality'' their \Imagen{} creations would have. For instance, P3{} quoted above added: ``Well, for me the image that it generated was sort of similar to what I envisioned in my mind... The way that it turned out is pretty cool. It is definitely not the style that I would have chosen for myself, my own drawings, but like it looks pretty.'' {(We should note that, due to the short term nature of our study, we are unable to study how such model steering impacts participant creativity over the long term.)}

Throughout these explorations, participants tended to improve or ``optimize'' a prompt if they found that at least one of the generated images was helpful. Participants refined their prompts both to bring out aspects they found successful in the initial set of results, and to steer results away from undesired properties. For example, while creating a Cheshire Cat, P9{} liked the \textit{card design} in the results, rather than the cat in the foreground:  ``[I] kind of like some of these designs..." They then updated their prompt to get more of that card design: \prompt{Frame with filigree pattern. Circus colors}. 

At other points, participants refined their prompts to steer results away from undesired properties. For example, P8{} first tried to generate a jovial Cheshire Cat image but remarked that “Those are a little terrifying." He thus updated the prompt to make the image look less scary and more festive: \prompt{Invitation to a birthday party. Alice in Wonderland. Cheshire Cat}. Similarly, participants noticed that with some images that were poorly cropped, they could obtain better results if they appended \prompt{framed painting} to their prompt.

\subsubsection{Novel images steered novel ideas}

Whereas \IS{} surfaced existing images on the Internet, \Imagen{} {allowed} participants to generate {entirely novel} images, {and allowed them to successfully explore their creative ideas}. For example, P5{} described hitting a wall with \IS{} when he could not find a specific aspect of what he wanted via \IS{}, possibly because it did not exist in the real world: ``I wanted a picture of a dolphin...And I started to Google it... One of my problems when I was searching around, is I couldn't quite get the image I wanted, right? I wanted to make something new and I couldn't quite get the right image I wanted."In contrast, participants were able to use \imagen{} to create novel combinations of ideas that did not exist, such as a giraffe driving a Lamborghini: ``Like `a giraffe is driving a Lamborghini' ... these are things you can never do. You can never have images, that look reasonable for those online. If you had to do it the old-fashioned way, or be really good at Photoshop or Illustrator. And it would take a lot longer than I have.'' They were also able to apply styles to content from a time period, which would not have been possible in the real world: ``...show me `the Apollo 11 Landing in the style of Dali'”.

\begin{figure}
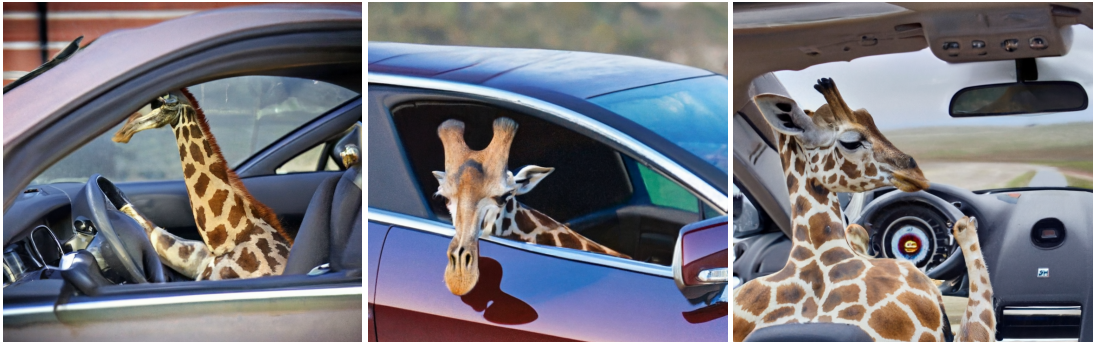

     \centering
     \begin{subfigure}[t]{0.3\textwidth}
         \centering
         \includegraphics[width=\textwidth]{\imagepath{Imagen_giraffe-driver}}
         \label{fig:image-search-1}
     \end{subfigure}
     \begin{subfigure}[t]{0.3\textwidth}
         \centering
         \includegraphics[width=\textwidth]{\imagepath{Imagen_giraffe-moping}}
         \label{fig:image-search-1}
     \end{subfigure}
     \begin{subfigure}[t]{0.3\textwidth}
         \centering
         \includegraphics[width=\textwidth]{\imagepath{Imagen_giraffe}}
         \label{fig:image-search-1}
     \end{subfigure}
     \caption{A few giraffes driving fancy cars. (Prompt inspired by participant: \prompt{Giraffe is driving a Lamborghini. f2.2}). By enabling the rapid realization of novel ideas and unlikely combinations, \LPIMs{} enable an  exploration of the design space and fluid collaboration.}
\end{figure}

{In addition to exploring existing creative ideas, surprising image results also spurred participants} to go in a different direction. For example, TTI  surprises inspired participants to consider aesthetic styles, compositions, or other design choices they hadn't initially considered: ``When I asked \imagen{} for a doodle of that \imagen{} blew me away with something that was a different art style than I imagined. That inspired me to seek out stuff in that same art style or to keep asking for doodles.” Together, participants saw \imagen{} as a {way to support their creativity in ways that were qualitatively different from previous tools}. As P6{} noted: ``it really kind of stirs, my creative juices or whatever whereas like Googling for images does not really stir that…''

However, \imagen{} also occasionally generated {non-sensical or clearly flawed} images, such as animals with incorrect anatomy or images of the Moon with two ``Earths'' in the background. Participants contrasted this with \IS{}, which offered more predictable results because they were authentic images from the Internet. This, in turn, allowed participants to {feel that the resulting images reasonably depicted what was in the images}  without closer scrutiny. For example, P5{} suggested:  “...I trusted the images that I look for are going to look somewhat more sane... I am not going to see half of two rabbits”. Participants also saw such predictability as necessary when looking for a specific image. For instance, one pair of designers used public-domain images of the first Moon Landing. For such images, correctness was crucial: “Most of the images we used are very specific. They are images that \Imagen{} cannot generate.”

\subsubsection{Designing with an opinionated model}\label{ssub:opinionated}
As \imagen{} would sometimes produce unexpected results, the participants often felt the need to guide or work around the model's limitations. 
For instance, P3{} noted their decision to use  the model to generate an image for the entire invitation from a single prompt rather than prompting for each part of the image and compositing them: “I suppose because we knew the limitations of \imagen{}, in terms of like, composing it for multiple images is, it sort of reduced what we could do with it.” Other participants were able to avoid design fixation, but with considerable effort. P5{} remarked: ``It felt like I was fighting it….I felt like it was helpful, but I also felt like I had to massage every word and select every character very carefully not to upset it so that it could generate something I wanted.'' Consistent with prior work, in these and other quotes, participants seemed to ascribe the role of an opinionated design partner to \imagen{} For instance, Koch et al. described how participants ascribed agency to the AI tool (with one participant even referring to it as "an eccentric collaborator") \cite{koch2020CollaborativeAI}. Similarly, in a study of an AI-based co-creation tool that generates sketches to inspire the user as they are actively sketching, users perceived the AI as a ``collaborative partner'' in the condition when the system communicated with them \cite{rezwana2022AICoCreation}. {(Because participants themselves anthropomorphized the model, we characterize it as opinionated, rather than using other terms, such as being biased.)}

While prompting {with this opinionated} model enabled participants to express high-level concepts at rapid speed, participants struggled to systematically control low-level details, such as position, layout, and which letters appear in the text (note, however, that some participants did use \Imagen{} to generate text in styles that Google Slides did not support, see Figure~f{fig:cake-choices}). For instance, P4{} wished for ``more controllability'': ``it is kind of agonizing to keep typing in very different versions of the same thing, and you are like, no, I just want his hand to be, like a little bit farther down.'' Similarly, P5{} expressed their frustration with how \Imagen{} sometimes cropped parts of an object in the image: ``So this is yeah, with this image, we can go outside the lines and get something that covers more of the screen while it is just focusing on the top hat...\textit{after a few moments}... I cannot.''

\subsection{How text-to-image models changed collaborative dynamics during design}
{Text-to-image models} modulated the collaborative practices among participants by creating new ways to fluidly combine ideas with prompts. At the same time, because prompts were so central to these collaborations, asymmetric access to the prompts changed collaborative roles and exploration.

\subsubsection{Prompts allow participants to fluidly combine ideas}
A core aspect of creative collaboration is the ability to combine, re-mix, and try out ideas from multiple people \cite{fauconnier2008way}. However, while using \IS{}, participants sometimes discovered that, even when they could agree and combine their ideas, those combinations of ideas were often hard to find within the search results. For instance, one participant noted: ``It was easy for us to sort of like agree and collaborate on ideas but then it was hard to find images that match those ideas.'' Similarly, during their design session, P4 said to their partner: ``I like the one that you had with the crazy paper vintage background,'' but later was unable to find images of candles in that preferred style: “something about beggars cannot be choosers.”

In contrast to \IS{}, with \imagen{}, participants were both able to combine ideas in their prompts and experiment rapidly with different ways of composing them together. For instance, in this conversation, P7{} fluidly added his ideas for fireworks to their prompt about rockets on the moon: 
\begin{description}
\item[P8{}] Another theme could be something to do with rockets.
\item[P7{}] Oh yeah, or like rocket fireworks. [Prompt: \prompt{Fireworks exploding in the shape of a space shuttle.}]
\end{description}

Furthermore, \imagen{} allowed participants to see a variety of \textit{generated} images and choose the ones that best matched their needs. Reacting to a set of \imagen{} results based on their partner's query, P6{} said: ``It [image on the left] does not get the idea of the party across. Let's go with the one on the right because it has like the  astronaut has a party hat.''

Finally, even though not this was not the focus of our study, participants often spoke about how they learned tricks for successful prompting socially. For instance, P5{} suggested how this process of social learning was fun: ``Like, it could be fun, especially when me and my coworkers are all sitting at my desk and people like, oh, take this [prompt] and see what it does.'' In many of our design sessions, we saw many such prompt modifications, such as using \prompt{...framed art}, or particular camera or lens types to mimic in the images generated, such as \prompt{...Sigma 85mm f1.4}. Once participants shared such prompt tricks with their partner, they often used them in their collaborative design work. 

\subsubsection{Asymmetric access to prompts, randomized generation,  hinder collaboration}
Often, the ability to iterate on a design was weighted towards whichever collaborator had access to the prompt, leading to asymmetric access. This was particularly prominent in situations where participants prototyped prompts in windows that were not shared with their partner, as in this exchange:

\begin{description}
\item[P1] \textit{(chuckles)} Okay, well I got something which will be sort of, kind of more appropriate maybe. So I'm gonna paste it here  
\item[P2]  \textit{(seeing the results)} Hey! That's pretty good. Okay….Yeah, even the Lander is partying! I think we go with this one.
\end{description}

In this situation, even though P1{} was able to share an exciting image result with P2{}, P2{} was not able to iterate on the design because he did not have access to the prompt. In this case, we noticed P2{} became increasingly reliant on P1{} to create images as the design session progressed. 

Even with access to prompts, generative models (including \imagen{}) typically use a random seed as input, so users see different image results on consecutive runs of the model, even when providing an identical prompt input. As a result, participants were sometimes unable to replicate previous results reliably, hampering collaboration.

\subsection{Prompts as reflective design material}
Throughout their design session with \IS{}, it seemed that participants merely saw \IS{} as a way to find the needed images. In contrast, when participants used \imagen{}, they displayed a nuanced, functional understanding of how prompts could be used to achieve their design objectives. Moreover, this understanding was not related to the technical aspects of how \imagen{} worked -- not once in our sessions or interviews did participants mention ``transformers'', ``diffusion models'', or even ``deep learning.'' Instead, they spoke about and enacted how  \imagen{} allowed them to rapidly explore a range of artistic possibilities and to collaborate.

These observations lead us to characterize prompts as \textit{design materials}. Below, we describe how participants exhibited a tacit understanding of \imagen{} and how prompts allowed for exploration and reflection on model actions (i.e., images generated) and reflection in action (i.e., through collaboratively editing the prompts).

Participants used and developed their tacit mental models of \Imagen{}'s design orientation throughout their design process. For instance, P3{} noticed how \Imagen{} framed the subjects in its images: ``Most of the images that get generated by \Imagen{} always push everything up to the front.'' Sometimes, participants tried to compensate for what they believed the model did not understand. P6{} noticed, for instance, ``It seems like it does not know what the Cheshire Cat is," changing their prompt from \prompt{An illustration of the Cheshire Cat from 'Alice in Wonderland'} to \prompt{An illustration of a cat with a large face smiling and looking at the camera}. Finally, participants sometimes generated images mostly to test what \Imagen{} might do with a prompt. For instance, P4{} said to their partner: “Oh, we could try that with like `1969 poster' or no… because the poster will make Imagen try to…? Let’s try that. `1969 poster'.” Then, examining the results, they decided: ``These are like the very artsy side which is probably less what we want. But they are still fun.”

\subsubsection{Prompts allow rapidly exploring the design space} 

In their role as reflective design materials, prompts allowed participants to rapidly explore their design's content, style, and layout. 
For instance, many participants opened multiple instances of \Imagen{} (in different browser tabs) to explore variations of a prompt, such as \prompt{Drawing of a Cheshire cat from Alice in Wonderland. Cartoon}, \prompt{Drawing of a Cheshire cat from Alice in Wonderland. Psychedelic}, and \prompt{Colorful drawing of a Cheshire cat from Alice in Wonderland. Cartoon.} 

Participants made these decisions with fluidity, interlaying decisions of content and style while navigating the limits of the model: 

\begin{description}
\item[P12{}] Yeah... it might be hard to get Alice eating cake. 

\item[P11{}] ...yeah 

\item[P12{}] maybe we could do something with like `the cake from Alice in Wonderland'.

\item[P11{}] Yeah. Yeah. Maybe if we can't get Alice's face out of it than we could use Alice's face... like a non copyrighted one from Google.

\item[P11{}] Yeah. And then we could probably just do like, a cake that says `eat me' on it, right? 

\item[P12{}] Hmm, yeah. You have like a style that we want for that?

\item[P11{}] It definitely, it's got to be like a cartoon one, at least. So we don't want it photorealistic.
\end{description}

Finally, because \imagen{} displayed multiple candidates per prompt, participants also could explore their design choices based on the results they obtained (see Figure~f{fig:cake-choices}). 

\begin{figure}[htb]
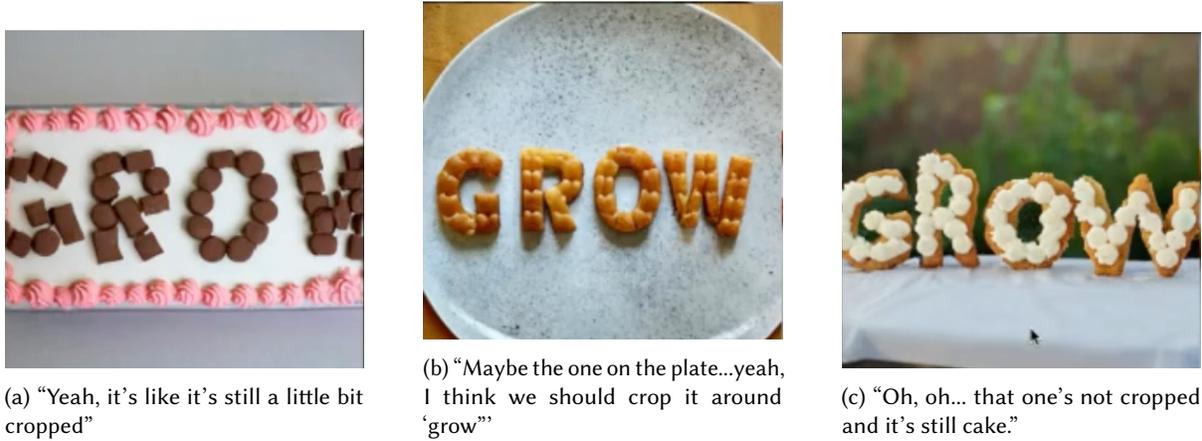

     \centering
     \begin{subfigure}[b]{0.3\textwidth}
         \centering
         \includegraphics[width=\textwidth]{\imagepath{grow-cake.png}}
         \caption{``Yeah, it's like it's still a little bit cropped''}
         \label{fig:grow-cake}
     \end{subfigure}
     \hfill
     \begin{subfigure}[b]{0.3\textwidth}
         \centering
         \includegraphics[width=\textwidth]{\imagepath{grow-on-plate.png}}
          \caption{``Maybe the one on the plate...yeah, I think we should crop it around `grow'''}
         \label{fig:cake2}
     \end{subfigure}
     \hfill
     \begin{subfigure}[b]{0.3\textwidth}
         \centering
         \includegraphics[width=\textwidth]{\imagepath{chosen-grow.png}}
          \caption{``Oh, oh...  that one's not cropped and it's still cake.''}
         \label{fig:cake3}
     \end{subfigure}
        \caption{\imagen{} allowed participants to rapidly explore the design space by allowing them to see different model interpretations of their prompt. Above, participant reactions to the prompt: \prompt{The word 'grow' made of cake.}}
        \label{fig:cake-choices}
\end{figure}

\subsubsection{A lack of distinction between means and ends}
A key distinction of robust design materials is their ability to merge ``means,'' and ``ends'' in the design process. As Schon writes, practice in such situations ``inquiry is not limited to a deliberation about means which depends on a prior agreement about the ends \cite{schon1987educating}. [They] do not keep means and ends separate, but define them interactively as they frame a problematic situation.'' Tacitly, perhaps, participants interactively and continuously framed their work throughout the design session.\\ 
\\

For instance, they often chose to dig deeper in the design space when it seemed promising. For instance: 
\begin{description}

\item[P7] \textit{(looking at screen)} Oh we're getting something! I will share this specific URL in the chat.
\item[P8] I like the fourth one.
\item[P7] Yeah. I will start adding some text if you want to keep iterating on this… I mean, I am OK, [if we] even replace the images that we have if we come up with something more party-like.
\end{description}

In other cases, prompts also allowed participants to discover new ``ends'' through other exploration-based ``means,'' e.g., P4{}: “I do not know. I just started trying to add stuff, but I agree. The ones we come up with since are better photos.” (This pair of participants replaced images in their final design.)

As noted elsewhere, reflective practice with prompts was far from perfect -- limited visibility of prompts between partners and an inability to replicate results even with the same prompt hampered collaboration and exploration. At the same time, framing prompts as design materials offer several opportunities to understand the model-aided design process better and build tools to improve it.

\subsection{Limitations.}
Some aspects of our study design complicate the interpretation of our findings. {We outline limitations here in three areas: participant composition, study design and analysis, and technology advancements}. 

Our participant pool was drawn among employees of one large US-based corporation, and does not cover the many possible ways that culture and training  might have shaped the design process with \LPIMs{}. {For example, they may be more comfortable collaborating remotely, as required in our study. Second, because of their choice to work in a technology firm,  it is very likely that they are more familiar with the idea of Artificial Intelligence than the general public. As a result, our findings likely are different from what might be expected with the general public. At the same time, as familiarity with AI grows in the future, it is possible that results with the general public are similar. At the same time, it is possible that our participants were more optimistic about the possibilities of technology, given their choice of employment. Because of company policies and laws, we were prevented from asking about sensitive demographic details such as race or national origin, and are unable to report differences among participants on these attributes, and if participants had differing concerns based on their identities. } 

{Our study focused on a single design task, which while representative of many tasks that non-professional designers engage in, may offer an incomplete picture of the impact of TTI models on design practices.} It was not possible to systematically observe every participant's prompt attempts, because some of those explorations were in screens that were not shared. Furthermore, our analysis is limited to observable conversations. For the interactions we \textit{could} observe, observing a designer's interactions with the model does not definitively indicate their conceptions; for example, designers who acted in similar ways even when they engaged different mental processes. Since our analysis was episodic rather than longitudinal, we are also unable to discover how design strategies  evolve within individual and pairs. 

{Finally, technological advances may lead to an evolution of some of our findings. After we conducted all our participant sessions, but before publication of this paper, new models such as Stable Diffusion were released, and led to advances like editing existing images (or parts of images) using textual prompts. Our conceptualization of prompts as design materials may extend to include these additional modalities, but future work should investigate specific ways in which such interactions influence exploration and reflection.}

\section{Discussion}\label{sec:discussion}
Our work contributes new insights into the use of \LPIMs{} in the design process by addressing our initial research questions:
\begin{itemize}
    \item RQ1: \textit{How does using prompt-based image generation change the design process?} Our qualitative results show that \LPIMs{} allow for rapid exploration of the design space, and designers use them as a new form of reflective design material by formulating descriptions of their image ideas, then coming up with scenarios for expressing, testing, and finally refining their prompts either by affirming their initial image ideas or formulating new ones.
    
    \item RQ2: \textit{How do prompt-based image generation models change collaborative dynamics during design?} Based on the pair collaborations we observed, prompt-based image generation also allows for fluid collaboration and rapid prototyping for the collaborative synthesis of ideas. However, prompt visibility and the non-determinism in current models modulated the effectiveness of such collaboration.
\end{itemize}

Informally, we also observed that participants were delighted both by the surprisingly good images and images that were ``hilariously bad''.
In addition, many participants reported that their experience with \imagen{} was more ``fun.'' (P4) Below, we discuss some emergent questions based on our findings. 

\subsection{Tool support for prompts as an indirect design material}
We found that the indirect nature of prompting both supported the design process (by augmenting creative freedom) and made it more challenging (while participants worked on rephrasing prompts to match intent). In some ways, prompts occupy a similar role in visual design as HTML did in early web design. By seeing how a webpage was constructed, designers could rapidly adopt good ideas, remix them, and popularize them widely. In such designerly practices, the role of Web browsers was also key -- by making ``View Source'' a universal feature, browsers likely transformed millions of people from web ``readers'' to web ``writers.'' Our work suggests that a similar ``View Source'' feature would also catalyze visual design. For instance, models could embed prompts as image EXIF data. Such tool support could also allow for more straightforward iteration and remixing and make prompt-based image generation even more accessible. For example, one could imagine mixing a partner's prompt (or prompt embedding) into one's prompt, creating a hybrid prompt. Finally, since prompts are text, many ideas from text editors and version control may also be relevant. For example, collaborating partners could send each other suggested edits, similar to text editors, or discover good prompting strategies (such as including clauses like \prompt{framed art}). Other materials such as tutorials or can also aid in this discovery (already, prompt guides are emerging).  

In addition, we can build on our observation that participants saw prompts as indirect design materials interpreted by opinionated image models. In light of these challenges, future work could empower end-users to control the model properties better. For example, users could swap out back-end models or fine-tune models on specific data (e.g., the same base model could have its weights fine-tuned by training on illustrations to help create art or furniture pictures for use by interior designers).

Low-level properties (e.g., position and layout) were challenging to control and remain an open question. Finally, while participants pushed back on opinionated models through prompt design, it may also be beneficial to develop models that are constrained in their output (e.g., not to crop parts of a salient object).

\subsection{Support for iterative design with prompts}
Prompt-based models are currently being optimized for "one-shot" interactions: each run of the model uses a random seed without the ability to be anchored on prior results. Interestingly, although models have been optimized for single-shot accuracy, users may be treating it more as a thinking tool, working incrementally, rephrasing, steering, and backtracking as a fundamental part of their design process.  

In the future, supporting \textit{iterative prompting} as a first-class objective would better allow people to use prompts as flexible design material. For example, users could toggle the random seed on and off depending on whether they were iterating on a design or starting in a new direction. Alternatively, they could bias the model to constrain or broaden exploration. Allowing users to navigate a history of states could also support exploring multiple ideas and backtracking. Since users may spend an inordinate amount of time rephrasing prompts to get the model output they desire, it may be valuable to combat potential design fixation by explicitly supporting the parallel exploration of \textit{multiple} ideas, a process that is known to improve design results \cite{dow2010parallel}. {A recent study showed that designers can directly interact with concepts in the model latent space by using semantic guidance to steer the diffusion process along several directions which enabled them to perform more sophisticated image composition and editing \cite{kwon2022diffusion}.}

\subsection{Multimodal affordances} 
The multi-modality of prompt-based models also creates new affordances. For example, such multimodality might help with greater control. In the current study, participants could not quickly iterate on a promising image result (e.g., ask the model to generate more results like Image 2, but with a bigger birthday cake) and instead resorted to updating the prompt with the desired property (e.g., giant birthday cake). Similarly, model support for multimodality might help with better composited images, for instance by moving an object into the background with pointer selection, and extending images for a wider field of view (with text prompts). Such affordances would improve iteration, and help designers express more complex visual concepts. 

\section{Conclusion}
This paper examines how \LPIMs{} can influence design processes and collaboration in goal-oriented design. Our results suggest that rather than a simple ``magic'' moment where designers input a prompt and designs are automatically generated, these models allow for a nuanced reflective practice of exploration, iteration, and collaboration. Our results also suggest that prompts can act as a design material and can support such emergent reflective practices. Finally, our study reveals several opportunities for tool design that build on the notion of prompts as reflective design materials, and suggest future directions for \LPIM{} research. Together, they point to a future where designers can use \LPIMs{} more effectively, resulting in a deeper and more creative practice.

\section{Acknowledgments}

Removed for anonymous review.

\bibliographystyle{ACM-Reference-Format}
\bibliography{sample-base, design, sections/inprogress}


\begin{thebibliography}{68}


\ifx \showCODEN    \undefined \def \showCODEN     #1{\unskip}     \fi
\ifx \showDOI      \undefined \def \showDOI       #1{#1}\fi
\ifx \showISBNx    \undefined \def \showISBNx     #1{\unskip}     \fi
\ifx \showISBNxiii \undefined \def \showISBNxiii  #1{\unskip}     \fi
\ifx \showISSN     \undefined \def \showISSN      #1{\unskip}     \fi
\ifx \showLCCN     \undefined \def \showLCCN      #1{\unskip}     \fi
\ifx \shownote     \undefined \def \shownote      #1{#1}          \fi
\ifx \showarticletitle \undefined \def \showarticletitle #1{#1}   \fi
\ifx \showURL      \undefined \def \showURL       {\relax}        \fi
\providecommand\bibfield[2]{#2}
\providecommand\bibinfo[2]{#2}
\providecommand\natexlab[1]{#1}
\providecommand\showeprint[2][]{arXiv:#2}

\bibitem[Dyn(2022)]%
        {Dynamicl51:online}
 \bibinfo{year}{2022}\natexlab{}.
\newblock \bibinfo{title}{Dynamicland}.
\newblock \bibinfo{howpublished}{\url{https://dynamicland.org/}}.
\newblock
\newblock
\shownote{(Accessed on 09/07/2022)}.


\bibitem[mid(2022)]%
        {midjourneyWeb}
 \bibinfo{year}{2022}\natexlab{}.
\newblock \bibinfo{title}{Midjourney}.
\newblock
\newblock
\urldef\tempurl%
\url{https://www.midjourney.com/home/}
\showURL{%
\tempurl}
\newblock
\shownote{(Accessed on 09/01/2022)}.


\bibitem[Scr(2022)]%
        {ScratchI75:online}
 \bibinfo{year}{2022}\natexlab{}.
\newblock \bibinfo{title}{Scratch - Imagine, Program, Share}.
\newblock \bibinfo{howpublished}{\url{https://scratch.mit.edu/}}.
\newblock
\newblock
\shownote{(Accessed on 09/07/2022)}.


\bibitem[AI(2022)]%
        {stableDiffusionWeb}
\bibfield{author}{\bibinfo{person}{Stability AI}.}
  \bibinfo{year}{2022}\natexlab{}.
\newblock \bibinfo{title}{Stable Diffusion launch announcement —
  Stability.Ai}.
\newblock
\newblock
\urldef\tempurl%
\url{https://stability.ai/blog/stable-diffusion-announcement}
\showURL{%
\tempurl}
\newblock
\shownote{(Accessed on 08/31/2022)}.


\bibitem[Betz et~al\mbox{.}(2021)]%
        {betz2021Aloud}
\bibfield{author}{\bibinfo{person}{Gregor Betz}, \bibinfo{person}{Kyle
  Richardson}, {and} \bibinfo{person}{Christian Voigt}.}
  \bibinfo{year}{2021}\natexlab{}.
\newblock \bibinfo{title}{Thinking Aloud: Dynamic Context Generation Improves
  Zero-Shot Reasoning Performance of GPT-2}.
\newblock
\newblock
\urldef\tempurl%
\url{https://doi.org/10.48550/ARXIV.2103.13033}
\showDOI{\tempurl}


\bibitem[Bommasani et~al\mbox{.}(2021)]%
        {bommasani2021Foundation}
\bibfield{author}{\bibinfo{person}{Rishi Bommasani}, \bibinfo{person}{Drew~A.
  Hudson}, \bibinfo{person}{Ehsan Adeli}, \bibinfo{person}{Russ Altman},
  \bibinfo{person}{Simran Arora}, \bibinfo{person}{Sydney von Arx},
  \bibinfo{person}{Michael~S. Bernstein}, \bibinfo{person}{Jeannette Bohg},
  \bibinfo{person}{Antoine Bosselut}, \bibinfo{person}{Emma Brunskill},
  \bibinfo{person}{Erik Brynjolfsson}, \bibinfo{person}{Shyamal Buch},
  \bibinfo{person}{Dallas Card}, \bibinfo{person}{Rodrigo Castellon},
  \bibinfo{person}{Niladri Chatterji}, \bibinfo{person}{Annie Chen},
  \bibinfo{person}{Kathleen Creel}, \bibinfo{person}{Jared~Quincy Davis},
  \bibinfo{person}{Dora Demszky}, \bibinfo{person}{Chris Donahue},
  \bibinfo{person}{Moussa Doumbouya}, \bibinfo{person}{Esin Durmus},
  \bibinfo{person}{Stefano Ermon}, \bibinfo{person}{John Etchemendy},
  \bibinfo{person}{Kawin Ethayarajh}, \bibinfo{person}{Li Fei-Fei},
  \bibinfo{person}{Chelsea Finn}, \bibinfo{person}{Trevor Gale},
  \bibinfo{person}{Lauren Gillespie}, \bibinfo{person}{Karan Goel},
  \bibinfo{person}{Noah Goodman}, \bibinfo{person}{Shelby Grossman},
  \bibinfo{person}{Neel Guha}, \bibinfo{person}{Tatsunori Hashimoto},
  \bibinfo{person}{Peter Henderson}, \bibinfo{person}{John Hewitt},
  \bibinfo{person}{Daniel~E. Ho}, \bibinfo{person}{Jenny Hong},
  \bibinfo{person}{Kyle Hsu}, \bibinfo{person}{Jing Huang},
  \bibinfo{person}{Thomas Icard}, \bibinfo{person}{Saahil Jain},
  \bibinfo{person}{Dan Jurafsky}, \bibinfo{person}{Pratyusha Kalluri},
  \bibinfo{person}{Siddharth Karamcheti}, \bibinfo{person}{Geoff Keeling},
  \bibinfo{person}{Fereshte Khani}, \bibinfo{person}{Omar Khattab},
  \bibinfo{person}{Pang~Wei Koh}, \bibinfo{person}{Mark Krass},
  \bibinfo{person}{Ranjay Krishna}, \bibinfo{person}{Rohith Kuditipudi},
  \bibinfo{person}{Ananya Kumar}, \bibinfo{person}{Faisal Ladhak},
  \bibinfo{person}{Mina Lee}, \bibinfo{person}{Tony Lee}, \bibinfo{person}{Jure
  Leskovec}, \bibinfo{person}{Isabelle Levent}, \bibinfo{person}{Xiang~Lisa
  Li}, \bibinfo{person}{Xuechen Li}, \bibinfo{person}{Tengyu Ma},
  \bibinfo{person}{Ali Malik}, \bibinfo{person}{Christopher~D. Manning},
  \bibinfo{person}{Suvir Mirchandani}, \bibinfo{person}{Eric Mitchell},
  \bibinfo{person}{Zanele Munyikwa}, \bibinfo{person}{Suraj Nair},
  \bibinfo{person}{Avanika Narayan}, \bibinfo{person}{Deepak Narayanan},
  \bibinfo{person}{Ben Newman}, \bibinfo{person}{Allen Nie},
  \bibinfo{person}{Juan~Carlos Niebles}, \bibinfo{person}{Hamed Nilforoshan},
  \bibinfo{person}{Julian Nyarko}, \bibinfo{person}{Giray Ogut},
  \bibinfo{person}{Laurel Orr}, \bibinfo{person}{Isabel Papadimitriou},
  \bibinfo{person}{Joon~Sung Park}, \bibinfo{person}{Chris Piech},
  \bibinfo{person}{Eva Portelance}, \bibinfo{person}{Christopher Potts},
  \bibinfo{person}{Aditi Raghunathan}, \bibinfo{person}{Rob Reich},
  \bibinfo{person}{Hongyu Ren}, \bibinfo{person}{Frieda Rong},
  \bibinfo{person}{Yusuf Roohani}, \bibinfo{person}{Camilo Ruiz},
  \bibinfo{person}{Jack Ryan}, \bibinfo{person}{Christopher Ré},
  \bibinfo{person}{Dorsa Sadigh}, \bibinfo{person}{Shiori Sagawa},
  \bibinfo{person}{Keshav Santhanam}, \bibinfo{person}{Andy Shih},
  \bibinfo{person}{Krishnan Srinivasan}, \bibinfo{person}{Alex Tamkin},
  \bibinfo{person}{Rohan Taori}, \bibinfo{person}{Armin~W. Thomas},
  \bibinfo{person}{Florian Tramèr}, \bibinfo{person}{Rose~E. Wang},
  \bibinfo{person}{William Wang}, \bibinfo{person}{Bohan Wu},
  \bibinfo{person}{Jiajun Wu}, \bibinfo{person}{Yuhuai Wu},
  \bibinfo{person}{Sang~Michael Xie}, \bibinfo{person}{Michihiro Yasunaga},
  \bibinfo{person}{Jiaxuan You}, \bibinfo{person}{Matei Zaharia},
  \bibinfo{person}{Michael Zhang}, \bibinfo{person}{Tianyi Zhang},
  \bibinfo{person}{Xikun Zhang}, \bibinfo{person}{Yuhui Zhang},
  \bibinfo{person}{Lucia Zheng}, \bibinfo{person}{Kaitlyn Zhou}, {and}
  \bibinfo{person}{Percy Liang}.} \bibinfo{year}{2021}\natexlab{}.
\newblock \bibinfo{title}{On the Opportunities and Risks of Foundation Models}.
\newblock
\newblock
\urldef\tempurl%
\url{https://doi.org/10.48550/ARXIV.2108.07258}
\showDOI{\tempurl}


\bibitem[Braun and Clarke(2006)]%
        {braun2006using}
\bibfield{author}{\bibinfo{person}{Virginia Braun} {and}
  \bibinfo{person}{Victoria Clarke}.} \bibinfo{year}{2006}\natexlab{}.
\newblock \showarticletitle{Using thematic analysis in psychology}.
\newblock \bibinfo{journal}{\emph{Qualitative research in psychology}}
  \bibinfo{volume}{3}, \bibinfo{number}{2} (\bibinfo{year}{2006}),
  \bibinfo{pages}{77--101}.
\newblock


\bibitem[Brock et~al\mbox{.}(2018)]%
        {gan2018Brock}
\bibfield{author}{\bibinfo{person}{Andrew Brock}, \bibinfo{person}{Jeff
  Donahue}, {and} \bibinfo{person}{Karen Simonyan}.}
  \bibinfo{year}{2018}\natexlab{}.
\newblock \showarticletitle{Large Scale {GAN} Training for High Fidelity
  Natural Image Synthesis}.
\newblock \bibinfo{journal}{\emph{CoRR}}  \bibinfo{volume}{abs/1809.11096}
  (\bibinfo{year}{2018}).
\newblock
\showeprint[arXiv]{1809.11096}
\urldef\tempurl%
\url{http://arxiv.org/abs/1809.11096}
\showURL{%
\tempurl}


\bibitem[Brown et~al\mbox{.}(2020)]%
        {NEURIPS2020_1457c0d6}
\bibfield{author}{\bibinfo{person}{Tom Brown}, \bibinfo{person}{Benjamin Mann},
  \bibinfo{person}{Nick Ryder}, \bibinfo{person}{Melanie Subbiah},
  \bibinfo{person}{Jared~D Kaplan}, \bibinfo{person}{Prafulla Dhariwal},
  \bibinfo{person}{Arvind Neelakantan}, \bibinfo{person}{Pranav Shyam},
  \bibinfo{person}{Girish Sastry}, \bibinfo{person}{Amanda Askell},
  \bibinfo{person}{Sandhini Agarwal}, \bibinfo{person}{Ariel Herbert-Voss},
  \bibinfo{person}{Gretchen Krueger}, \bibinfo{person}{Tom Henighan},
  \bibinfo{person}{Rewon Child}, \bibinfo{person}{Aditya Ramesh},
  \bibinfo{person}{Daniel Ziegler}, \bibinfo{person}{Jeffrey Wu},
  \bibinfo{person}{Clemens Winter}, \bibinfo{person}{Chris Hesse},
  \bibinfo{person}{Mark Chen}, \bibinfo{person}{Eric Sigler},
  \bibinfo{person}{Mateusz Litwin}, \bibinfo{person}{Scott Gray},
  \bibinfo{person}{Benjamin Chess}, \bibinfo{person}{Jack Clark},
  \bibinfo{person}{Christopher Berner}, \bibinfo{person}{Sam McCandlish},
  \bibinfo{person}{Alec Radford}, \bibinfo{person}{Ilya Sutskever}, {and}
  \bibinfo{person}{Dario Amodei}.} \bibinfo{year}{2020}\natexlab{}.
\newblock \showarticletitle{Language Models are Few-Shot Learners}. In
  \bibinfo{booktitle}{\emph{Advances in Neural Information Processing
  Systems}}, \bibfield{editor}{\bibinfo{person}{H.~Larochelle},
  \bibinfo{person}{M.~Ranzato}, \bibinfo{person}{R.~Hadsell},
  \bibinfo{person}{M.F. Balcan}, {and} \bibinfo{person}{H.~Lin}} (Eds.),
  Vol.~\bibinfo{volume}{33}. \bibinfo{publisher}{Curran Associates, Inc.},
  \bibinfo{pages}{1877--1901}.
\newblock
\urldef\tempurl%
\url{https://proceedings.neurips.cc/paper/2020/file/1457c0d6bfcb4967418bfb8ac142f64a-Paper.pdf}
\showURL{%
\tempurl}


\bibitem[Bruckman(1998)]%
        {bruckman1998community}
\bibfield{author}{\bibinfo{person}{Amy Bruckman}.}
  \bibinfo{year}{1998}\natexlab{}.
\newblock \showarticletitle{Community support for constructionist learning}.
\newblock \bibinfo{journal}{\emph{Computer Supported Cooperative Work (CSCW)}}
  \bibinfo{volume}{7}, \bibinfo{number}{1} (\bibinfo{year}{1998}),
  \bibinfo{pages}{47--86}.
\newblock


\bibitem[Buschek et~al\mbox{.}(2021)]%
        {buschek2021nine}
\bibfield{author}{\bibinfo{person}{Daniel Buschek}, \bibinfo{person}{Lukas
  Mecke}, \bibinfo{person}{Florian Lehmann}, {and} \bibinfo{person}{Hai Dang}.}
  \bibinfo{year}{2021}\natexlab{}.
\newblock \showarticletitle{Nine Potential Pitfalls when Designing Human-AI
  Co-Creative Systems}.
\newblock \bibinfo{journal}{\emph{Workshops at the International Conference on
  Intelligent User Interfaces (IUI)}} (\bibinfo{year}{2021}).
\newblock


\bibitem[Clark et~al\mbox{.}(2018)]%
        {clark2018Writing}
\bibfield{author}{\bibinfo{person}{Elizabeth Clark},
  \bibinfo{person}{Anne~Spencer Ross}, \bibinfo{person}{Chenhao Tan},
  \bibinfo{person}{Yangfeng Ji}, {and} \bibinfo{person}{Noah~A. Smith}.}
  \bibinfo{year}{2018}\natexlab{}.
\newblock \showarticletitle{Creative Writing with a Machine in the Loop: Case
  Studies on Slogans and Stories}. In \bibinfo{booktitle}{\emph{23rd
  International Conference on Intelligent User Interfaces}} (Tokyo, Japan)
  \emph{(\bibinfo{series}{IUI '18})}. \bibinfo{publisher}{Association for
  Computing Machinery}, \bibinfo{address}{New York, NY, USA},
  \bibinfo{pages}{329–340}.
\newblock
\showISBNx{9781450349451}
\urldef\tempurl%
\url{https://doi.org/10.1145/3172944.3172983}
\showDOI{\tempurl}


\bibitem[Dasgupta et~al\mbox{.}(2016)]%
        {dasgupta2016remixing}
\bibfield{author}{\bibinfo{person}{Sayamindu Dasgupta},
  \bibinfo{person}{William Hale}, \bibinfo{person}{Andr{\'e}s
  Monroy-Hern{\'a}ndez}, {and} \bibinfo{person}{Benjamin~Mako Hill}.}
  \bibinfo{year}{2016}\natexlab{}.
\newblock \showarticletitle{Remixing as a pathway to computational thinking}.
  In \bibinfo{booktitle}{\emph{Proceedings of the 19th ACM Conference on
  Computer-Supported Cooperative Work \& Social Computing}}.
  \bibinfo{pages}{1438--1449}.
\newblock


\bibitem[Davis et~al\mbox{.}(2015)]%
        {davis2015Drawing}
\bibfield{author}{\bibinfo{person}{Nicholas Davis}, \bibinfo{person}{Chih-PIn
  Hsiao}, \bibinfo{person}{Kunwar~Yashraj Singh}, \bibinfo{person}{Lisa Li},
  \bibinfo{person}{Sanat Moningi}, {and} \bibinfo{person}{Brian Magerko}.}
  \bibinfo{year}{2015}\natexlab{}.
\newblock \showarticletitle{Drawing Apprentice: An Enactive Co-Creative Agent
  for Artistic Collaboration}. In \bibinfo{booktitle}{\emph{Proceedings of the
  2015 ACM SIGCHI Conference on Creativity and Cognition}} (Glasgow, United
  Kingdom) \emph{(\bibinfo{series}{C\&C '15})}. \bibinfo{publisher}{Association
  for Computing Machinery}, \bibinfo{address}{New York, NY, USA},
  \bibinfo{pages}{185–186}.
\newblock
\showISBNx{9781450335980}
\urldef\tempurl%
\url{https://doi.org/10.1145/2757226.2764555}
\showDOI{\tempurl}


\bibitem[Davis et~al\mbox{.}(2016)]%
        {davis2016Drawing}
\bibfield{author}{\bibinfo{person}{Nicholas Davis}, \bibinfo{person}{Chih-PIn
  Hsiao}, \bibinfo{person}{Kunwar Yashraj~Singh}, \bibinfo{person}{Lisa Li},
  {and} \bibinfo{person}{Brian Magerko}.} \bibinfo{year}{2016}\natexlab{}.
\newblock \showarticletitle{Empirically Studying Participatory Sense-Making in
  Abstract Drawing with a Co-Creative Cognitive Agent}. In
  \bibinfo{booktitle}{\emph{Proceedings of the 21st International Conference on
  Intelligent User Interfaces}} (Sonoma, California, USA)
  \emph{(\bibinfo{series}{IUI '16})}. \bibinfo{publisher}{Association for
  Computing Machinery}, \bibinfo{address}{New York, NY, USA},
  \bibinfo{pages}{196–207}.
\newblock
\showISBNx{9781450341370}
\urldef\tempurl%
\url{https://doi.org/10.1145/2856767.2856795}
\showDOI{\tempurl}


\bibitem[Dow et~al\mbox{.}(2010)]%
        {dow2010parallel}
\bibfield{author}{\bibinfo{person}{Steven~P Dow}, \bibinfo{person}{Alana
  Glassco}, \bibinfo{person}{Jonathan Kass}, \bibinfo{person}{Melissa Schwarz},
  \bibinfo{person}{Daniel~L Schwartz}, {and} \bibinfo{person}{Scott~R
  Klemmer}.} \bibinfo{year}{2010}\natexlab{}.
\newblock \showarticletitle{Parallel prototyping leads to better design
  results, more divergence, and increased self-efficacy}.
\newblock \bibinfo{journal}{\emph{ACM Transactions on Computer-Human
  Interaction (TOCHI)}} \bibinfo{volume}{17}, \bibinfo{number}{4}
  (\bibinfo{year}{2010}), \bibinfo{pages}{1--24}.
\newblock


\bibitem[Edwards(2022)]%
        {AIwinsst11:online}
\bibfield{author}{\bibinfo{person}{Benj Edwards}.}
  \bibinfo{year}{2022}\natexlab{}.
\newblock \bibinfo{title}{AI wins state fair art contest, annoys humans}.
\newblock
\newblock
\urldef\tempurl%
\url{https://arstechnica.com/information-technology/2022/08/ai-wins-state-fair-art-contest-annoys-humans/}
\showURL{%
\tempurl}
\newblock
\shownote{(Accessed on 09/02/2022)}.


\bibitem[Fauconnier and Turner(2008)]%
        {fauconnier2008way}
\bibfield{author}{\bibinfo{person}{Gilles Fauconnier} {and}
  \bibinfo{person}{Mark Turner}.} \bibinfo{year}{2008}\natexlab{}.
\newblock \bibinfo{booktitle}{\emph{The way we think: Conceptual blending and
  the mind's hidden complexities}}.
\newblock \bibinfo{publisher}{Basic books}.
\newblock


\bibitem[Frich et~al\mbox{.}(2019)]%
        {frih2019CSTinHCI}
\bibfield{author}{\bibinfo{person}{Jonas Frich}, \bibinfo{person}{Lindsay
  MacDonald~Vermeulen}, \bibinfo{person}{Christian Remy},
  \bibinfo{person}{Michael~Mose Biskjaer}, {and} \bibinfo{person}{Peter
  Dalsgaard}.} \bibinfo{year}{2019}\natexlab{}.
\newblock \showarticletitle{Mapping the Landscape of Creativity Support Tools
  in HCI}. In \bibinfo{booktitle}{\emph{Proceedings of the 2019 CHI Conference
  on Human Factors in Computing Systems}} (Glasgow, Scotland Uk)
  \emph{(\bibinfo{series}{CHI '19})}. \bibinfo{publisher}{Association for
  Computing Machinery}, \bibinfo{address}{New York, NY, USA},
  \bibinfo{pages}{1–18}.
\newblock
\showISBNx{9781450359702}
\urldef\tempurl%
\url{https://doi.org/10.1145/3290605.3300619}
\showDOI{\tempurl}


\bibitem[Gero and Chilton(2019)]%
        {gero2019Metaphoria}
\bibfield{author}{\bibinfo{person}{Katy~Ilonka Gero} {and}
  \bibinfo{person}{Lydia~B. Chilton}.} \bibinfo{year}{2019}\natexlab{}.
\newblock \showarticletitle{Metaphoria: An Algorithmic Companion for Metaphor
  Creation}. In \bibinfo{booktitle}{\emph{Proceedings of the 2019 CHI
  Conference on Human Factors in Computing Systems}} (Glasgow, Scotland Uk)
  \emph{(\bibinfo{series}{CHI '19})}. \bibinfo{publisher}{Association for
  Computing Machinery}, \bibinfo{address}{New York, NY, USA},
  \bibinfo{pages}{1–12}.
\newblock
\showISBNx{9781450359702}
\urldef\tempurl%
\url{https://doi.org/10.1145/3290605.3300526}
\showDOI{\tempurl}


\bibitem[Ghajargar and Wiberg(2018)]%
        {ghajargar2018thinking}
\bibfield{author}{\bibinfo{person}{Maliheh Ghajargar} {and}
  \bibinfo{person}{Mikael Wiberg}.} \bibinfo{year}{2018}\natexlab{}.
\newblock \showarticletitle{Thinking with interactive artifacts: Reflection as
  a concept in design outcomes}.
\newblock \bibinfo{journal}{\emph{Design Issues}} \bibinfo{volume}{34},
  \bibinfo{number}{2} (\bibinfo{year}{2018}), \bibinfo{pages}{48--63}.
\newblock


\bibitem[Goodfellow et~al\mbox{.}(2020)]%
        {gan2014Goodfellow}
\bibfield{author}{\bibinfo{person}{Ian Goodfellow}, \bibinfo{person}{Jean
  Pouget-Abadie}, \bibinfo{person}{Mehdi Mirza}, \bibinfo{person}{Bing Xu},
  \bibinfo{person}{David Warde-Farley}, \bibinfo{person}{Sherjil Ozair},
  \bibinfo{person}{Aaron Courville}, {and} \bibinfo{person}{Yoshua Bengio}.}
  \bibinfo{year}{2020}\natexlab{}.
\newblock \showarticletitle{Generative Adversarial Networks}.
\newblock \bibinfo{journal}{\emph{Commun. ACM}} \bibinfo{volume}{63},
  \bibinfo{number}{11} (\bibinfo{date}{oct} \bibinfo{year}{2020}),
  \bibinfo{pages}{139–144}.
\newblock
\showISSN{0001-0782}
\urldef\tempurl%
\url{https://doi.org/10.1145/3422622}
\showDOI{\tempurl}


\bibitem[Google(2022a)]%
        {imagenWeb}
\bibfield{author}{\bibinfo{person}{Google}.} \bibinfo{year}{2022}\natexlab{a}.
\newblock \bibinfo{title}{Imagen: Text-to-Image Diffusion Models}.
\newblock
\newblock
\urldef\tempurl%
\url{https://imagen.research.google/}
\showURL{%
\tempurl}
\newblock
\shownote{(Accessed on 08/31/2022)}.


\bibitem[Google(2022b)]%
        {partiWeb}
\bibfield{author}{\bibinfo{person}{Google}.} \bibinfo{year}{2022}\natexlab{b}.
\newblock \bibinfo{title}{Parti: Pathways Autoregressive Text-to-Image Model}.
\newblock
\newblock
\urldef\tempurl%
\url{https://parti.research.google/}
\showURL{%
\tempurl}
\newblock
\shownote{(Accessed on 08/31/2022)}.


\bibitem[Hancock et~al\mbox{.}(2020)]%
        {hancock2020AICommunication}
\bibfield{author}{\bibinfo{person}{Jeffrey~T Hancock}, \bibinfo{person}{Mor
  Naaman}, {and} \bibinfo{person}{Karen Levy}.}
  \bibinfo{year}{2020}\natexlab{}.
\newblock \showarticletitle{{AI-Mediated Communication: Definition, Research
  Agenda, and Ethical Considerations}}.
\newblock \bibinfo{journal}{\emph{Journal of Computer-Mediated Communication}}
  \bibinfo{volume}{25}, \bibinfo{number}{1} (\bibinfo{date}{01}
  \bibinfo{year}{2020}), \bibinfo{pages}{89--100}.
\newblock
\showISSN{1083-6101}
\urldef\tempurl%
\url{https://doi.org/10.1093/jcmc/zmz022}
\showDOI{\tempurl}
\showeprint{https://academic.oup.com/jcmc/article-pdf/25/1/89/32961176/zmz022.pdf}


\bibitem[Huang et~al\mbox{.}(2020)]%
        {huang2020AISong}
\bibfield{author}{\bibinfo{person}{Cheng-Zhi~Anna Huang},
  \bibinfo{person}{Hendrik~Vincent Koops}, \bibinfo{person}{Ed Newton-Rex},
  \bibinfo{person}{Monica Dinculescu}, {and} \bibinfo{person}{Carrie~J. Cai}.}
  \bibinfo{year}{2020}\natexlab{}.
\newblock \showarticletitle{AI Song Contest: Human-AI Co-Creation in
  Songwriting}.
\newblock  (\bibinfo{year}{2020}).
\newblock
\urldef\tempurl%
\url{https://doi.org/10.48550/ARXIV.2010.05388}
\showDOI{\tempurl}


\bibitem[Jehn and Mannix(2001)]%
        {jehn2001dynamic}
\bibfield{author}{\bibinfo{person}{Karen~A Jehn} {and}
  \bibinfo{person}{Elizabeth~A Mannix}.} \bibinfo{year}{2001}\natexlab{}.
\newblock \showarticletitle{The dynamic nature of conflict: A longitudinal
  study of intragroup conflict and group performance}.
\newblock \bibinfo{journal}{\emph{Academy of management journal}}
  \bibinfo{volume}{44}, \bibinfo{number}{2} (\bibinfo{year}{2001}),
  \bibinfo{pages}{238--251}.
\newblock


\bibitem[Jeon et~al\mbox{.}(2021)]%
        {jeon2021Fashion}
\bibfield{author}{\bibinfo{person}{Youngseung Jeon}, \bibinfo{person}{Seungwan
  Jin}, \bibinfo{person}{Patrick~C. Shih}, {and} \bibinfo{person}{Kyungsik
  Han}.} \bibinfo{year}{2021}\natexlab{}.
\newblock \showarticletitle{FashionQ: An AI-Driven Creativity Support Tool for
  Facilitating Ideation in Fashion Design}. In
  \bibinfo{booktitle}{\emph{Proceedings of the 2021 CHI Conference on Human
  Factors in Computing Systems}} (Yokohama, Japan) \emph{(\bibinfo{series}{CHI
  '21})}. \bibinfo{publisher}{Association for Computing Machinery},
  \bibinfo{address}{New York, NY, USA}, Article \bibinfo{articleno}{576},
  \bibinfo{numpages}{18}~pages.
\newblock
\showISBNx{9781450380966}
\urldef\tempurl%
\url{https://doi.org/10.1145/3411764.3445093}
\showDOI{\tempurl}


\bibitem[Kantosalo and Toivonen(2016)]%
        {kantosalo2016modes}
\bibfield{author}{\bibinfo{person}{Anna Kantosalo} {and} \bibinfo{person}{Hannu
  Toivonen}.} \bibinfo{year}{2016}\natexlab{}.
\newblock \showarticletitle{Modes for creative human-computer collaboration:
  Alternating and task-divided co-creativity}. In
  \bibinfo{booktitle}{\emph{Proceedings of the seventh international conference
  on computational creativity}}. \bibinfo{pages}{77--84}.
\newblock


\bibitem[Karimi et~al\mbox{.}(2019)]%
        {karimi2019Drawing}
\bibfield{author}{\bibinfo{person}{Pegah Karimi}, \bibinfo{person}{Nicholas
  Davis}, \bibinfo{person}{Mary~Lou Maher}, \bibinfo{person}{Kazjon Grace},
  {and} \bibinfo{person}{Lina Lee}.} \bibinfo{year}{2019}\natexlab{}.
\newblock \showarticletitle{Relating Cognitive Models of Design Creativity to
  the Similarity of Sketches Generated by an AI Partner}. In
  \bibinfo{booktitle}{\emph{Proceedings of the 2019 on Creativity and
  Cognition}} (San Diego, CA, USA) \emph{(\bibinfo{series}{C\&C '19})}.
  \bibinfo{publisher}{Association for Computing Machinery},
  \bibinfo{address}{New York, NY, USA}, \bibinfo{pages}{259–270}.
\newblock
\showISBNx{9781450359177}
\urldef\tempurl%
\url{https://doi.org/10.1145/3325480.3325488}
\showDOI{\tempurl}


\bibitem[Karras et~al\mbox{.}(2019)]%
        {Karras_2019_CVPR}
\bibfield{author}{\bibinfo{person}{Tero Karras}, \bibinfo{person}{Samuli
  Laine}, {and} \bibinfo{person}{Timo Aila}.} \bibinfo{year}{2019}\natexlab{}.
\newblock \showarticletitle{A Style-Based Generator Architecture for Generative
  Adversarial Networks}. In \bibinfo{booktitle}{\emph{Proceedings of the
  IEEE/CVF Conference on Computer Vision and Pattern Recognition (CVPR)}}.
\newblock


\bibitem[Koch et~al\mbox{.}(2019)]%
        {koch2019Ideation}
\bibfield{author}{\bibinfo{person}{Janin Koch}, \bibinfo{person}{Andr\'{e}s
  Lucero}, \bibinfo{person}{Lena Hegemann}, {and} \bibinfo{person}{Antti
  Oulasvirta}.} \bibinfo{year}{2019}\natexlab{}.
\newblock \showarticletitle{May AI? Design Ideation with Cooperative Contextual
  Bandits}. In \bibinfo{booktitle}{\emph{Proceedings of the 2019 CHI Conference
  on Human Factors in Computing Systems}} (Glasgow, Scotland Uk)
  \emph{(\bibinfo{series}{CHI '19})}. \bibinfo{publisher}{Association for
  Computing Machinery}, \bibinfo{address}{New York, NY, USA},
  \bibinfo{pages}{1–12}.
\newblock
\showISBNx{9781450359702}
\urldef\tempurl%
\url{https://doi.org/10.1145/3290605.3300863}
\showDOI{\tempurl}


\bibitem[Koch et~al\mbox{.}(2020)]%
        {koch2020CollaborativeAI}
\bibfield{author}{\bibinfo{person}{Janin Koch}, \bibinfo{person}{Nicolas
  Taffin}, \bibinfo{person}{Michel Beaudouin-Lafon}, \bibinfo{person}{Markku
  Laine}, \bibinfo{person}{Andr\'{e}s Lucero}, {and} \bibinfo{person}{Wendy~E.
  Mackay}.} \bibinfo{year}{2020}\natexlab{}.
\newblock \showarticletitle{ImageSense: An Intelligent Collaborative Ideation
  Tool to Support Diverse Human-Computer Partnerships}.
\newblock \bibinfo{journal}{\emph{Proc. ACM Hum.-Comput. Interact.}}
  \bibinfo{volume}{4}, \bibinfo{number}{CSCW1}, Article \bibinfo{articleno}{45}
  (\bibinfo{date}{may} \bibinfo{year}{2020}), \bibinfo{numpages}{27}~pages.
\newblock
\urldef\tempurl%
\url{https://doi.org/10.1145/3392850}
\showDOI{\tempurl}


\bibitem[Kwon et~al\mbox{.}(2022)]%
        {kwon2022diffusion}
\bibfield{author}{\bibinfo{person}{Mingi Kwon}, \bibinfo{person}{Jaeseok
  Jeong}, {and} \bibinfo{person}{Youngjung Uh}.}
  \bibinfo{year}{2022}\natexlab{}.
\newblock \showarticletitle{Diffusion models already have a semantic latent
  space}.
\newblock \bibinfo{journal}{\emph{arXiv preprint arXiv:2210.10960}}
  (\bibinfo{year}{2022}).
\newblock


\bibitem[Lieber et~al\mbox{.}(2021)]%
        {lieberjurassic}
\bibfield{author}{\bibinfo{person}{O Lieber}, \bibinfo{person}{O Sharir},
  \bibinfo{person}{B Lentz}, {and} \bibinfo{person}{Y Shoham}.}
  \bibinfo{year}{2021}\natexlab{}.
\newblock \bibinfo{title}{Jurassic-1: Technical Details and Evaluation, White
  paper, AI21 Labs, 2021}.
\newblock
\newblock
\urldef\tempurl%
\url{URL:
  https://uploads-ssl.webflow.com/60fd4503684b466578c0d307/61138924626a6981ee09caf6\_jurassic\_tech\_paper.pdf}
\showURL{%
\tempurl}


\bibitem[Liu et~al\mbox{.}(2021)]%
        {liu2021InContext}
\bibfield{author}{\bibinfo{person}{Jiachang Liu}, \bibinfo{person}{Dinghan
  Shen}, \bibinfo{person}{Yizhe Zhang}, \bibinfo{person}{Bill Dolan},
  \bibinfo{person}{Lawrence Carin}, {and} \bibinfo{person}{Weizhu Chen}.}
  \bibinfo{year}{2021}\natexlab{}.
\newblock \bibinfo{title}{What Makes Good In-Context Examples for GPT-$3$?}
\newblock
\newblock
\urldef\tempurl%
\url{https://doi.org/10.48550/ARXIV.2101.06804}
\showDOI{\tempurl}


\bibitem[Louie et~al\mbox{.}(2020)]%
        {louie2020Music}
\bibfield{author}{\bibinfo{person}{Ryan Louie}, \bibinfo{person}{Andy Coenen},
  \bibinfo{person}{Cheng~Zhi Huang}, \bibinfo{person}{Michael Terry}, {and}
  \bibinfo{person}{Carrie~J. Cai}.} \bibinfo{year}{2020}\natexlab{}.
\newblock \showarticletitle{Novice-AI Music Co-Creation via AI-Steering Tools
  for Deep Generative Models}. In \bibinfo{booktitle}{\emph{Proceedings of the
  2020 CHI Conference on Human Factors in Computing Systems}} (Honolulu, HI,
  USA) \emph{(\bibinfo{series}{CHI '20})}. \bibinfo{publisher}{Association for
  Computing Machinery}, \bibinfo{address}{New York, NY, USA},
  \bibinfo{pages}{1–13}.
\newblock
\showISBNx{9781450367080}
\urldef\tempurl%
\url{https://doi.org/10.1145/3313831.3376739}
\showDOI{\tempurl}


\bibitem[Lu et~al\mbox{.}(2021)]%
        {lu2021PromptOrder}
\bibfield{author}{\bibinfo{person}{Yao Lu}, \bibinfo{person}{Max Bartolo},
  \bibinfo{person}{Alastair Moore}, \bibinfo{person}{Sebastian Riedel}, {and}
  \bibinfo{person}{Pontus Stenetorp}.} \bibinfo{year}{2021}\natexlab{}.
\newblock \bibinfo{title}{Fantastically Ordered Prompts and Where to Find Them:
  Overcoming Few-Shot Prompt Order Sensitivity}.
\newblock
\newblock
\urldef\tempurl%
\url{https://doi.org/10.48550/ARXIV.2104.08786}
\showDOI{\tempurl}


\bibitem[Mansimov et~al\mbox{.}(2015)]%
        {mansimov2015ImgFromCaption}
\bibfield{author}{\bibinfo{person}{Elman Mansimov}, \bibinfo{person}{Emilio
  Parisotto}, \bibinfo{person}{Jimmy~Lei Ba}, {and} \bibinfo{person}{Ruslan
  Salakhutdinov}.} \bibinfo{year}{2015}\natexlab{}.
\newblock \bibinfo{title}{Generating Images from Captions with Attention}.
\newblock
\newblock
\urldef\tempurl%
\url{https://doi.org/10.48550/ARXIV.1511.02793}
\showDOI{\tempurl}


\bibitem[Marlow and Dabbish(2014)]%
        {marlow2014rookie}
\bibfield{author}{\bibinfo{person}{Jennifer Marlow} {and}
  \bibinfo{person}{Laura Dabbish}.} \bibinfo{year}{2014}\natexlab{}.
\newblock \showarticletitle{From rookie to all-star: professional development
  in a graphic design social networking site}. In
  \bibinfo{booktitle}{\emph{Proceedings of the 17th ACM conference on Computer
  supported cooperative work \& social computing}}. \bibinfo{pages}{922--933}.
\newblock


\bibitem[McCormack et~al\mbox{.}(2019)]%
        {mccormack2019Music}
\bibfield{author}{\bibinfo{person}{Jon McCormack}, \bibinfo{person}{Toby
  Gifford}, \bibinfo{person}{Patrick Hutchings}, \bibinfo{person}{Maria~Teresa
  Llano~Rodriguez}, \bibinfo{person}{Matthew Yee-King}, {and}
  \bibinfo{person}{Mark d'Inverno}.} \bibinfo{year}{2019}\natexlab{}.
\newblock \showarticletitle{In a Silent Way: Communication Between AI and
  Improvising Musicians Beyond Sound}. In \bibinfo{booktitle}{\emph{Proceedings
  of the 2019 CHI Conference on Human Factors in Computing Systems}} (Glasgow,
  Scotland Uk) \emph{(\bibinfo{series}{CHI '19})}.
  \bibinfo{publisher}{Association for Computing Machinery},
  \bibinfo{address}{New York, NY, USA}, \bibinfo{pages}{1–11}.
\newblock
\showISBNx{9781450359702}
\urldef\tempurl%
\url{https://doi.org/10.1145/3290605.3300268}
\showDOI{\tempurl}


\bibitem[Miles and Huberman(1984)]%
        {miles1984drawing}
\bibfield{author}{\bibinfo{person}{Matthew~B Miles} {and}
  \bibinfo{person}{A~Michael Huberman}.} \bibinfo{year}{1984}\natexlab{}.
\newblock \showarticletitle{Drawing valid meaning from qualitative data: Toward
  a shared craft}.
\newblock \bibinfo{journal}{\emph{Educational researcher}}
  \bibinfo{volume}{13}, \bibinfo{number}{5} (\bibinfo{year}{1984}),
  \bibinfo{pages}{20--30}.
\newblock


\bibitem[Oh et~al\mbox{.}(2018)]%
        {oh2018Drawing}
\bibfield{author}{\bibinfo{person}{Changhoon Oh}, \bibinfo{person}{Jungwoo
  Song}, \bibinfo{person}{Jinhan Choi}, \bibinfo{person}{Seonghyeon Kim},
  \bibinfo{person}{Sungwoo Lee}, {and} \bibinfo{person}{Bongwon Suh}.}
  \bibinfo{year}{2018}\natexlab{}.
\newblock \showarticletitle{I Lead, You Help but Only with Enough Details:
  Understanding User Experience of Co-Creation with Artificial Intelligence}.
  In \bibinfo{booktitle}{\emph{Proceedings of the 2018 CHI Conference on Human
  Factors in Computing Systems}} (Montreal QC, Canada)
  \emph{(\bibinfo{series}{CHI '18})}. \bibinfo{publisher}{Association for
  Computing Machinery}, \bibinfo{address}{New York, NY, USA},
  \bibinfo{pages}{1–13}.
\newblock
\showISBNx{9781450356206}
\urldef\tempurl%
\url{https://doi.org/10.1145/3173574.3174223}
\showDOI{\tempurl}


\bibitem[OpenAI(2022a)]%
        {dalle2Web}
\bibfield{author}{\bibinfo{person}{OpenAI}.} \bibinfo{year}{2022}\natexlab{a}.
\newblock \bibinfo{title}{DALL·E 2}.
\newblock
\newblock
\urldef\tempurl%
\url{https://openai.com/dall-e-2/}
\showURL{%
\tempurl}
\newblock
\shownote{(Accessed on 08/31/2022)}.


\bibitem[OpenAI(2022b)]%
        {dalle1Web}
\bibfield{author}{\bibinfo{person}{OpenAI}.} \bibinfo{year}{2022}\natexlab{b}.
\newblock \bibinfo{title}{DALL·E: Creating Images from Text}.
\newblock
\newblock
\urldef\tempurl%
\url{https://openai.com/blog/dall-e/}
\showURL{%
\tempurl}
\newblock
\shownote{(Accessed on 08/31/2022)}.


\bibitem[Oviatt and Cohen(2000)]%
        {oviatt2000multimodalnatural}
\bibfield{author}{\bibinfo{person}{Sharon Oviatt} {and} \bibinfo{person}{Philip
  Cohen}.} \bibinfo{year}{2000}\natexlab{}.
\newblock \showarticletitle{Perceptual User Interfaces: Multimodal Interfaces
  That Process What Comes Naturally}.
\newblock \bibinfo{journal}{\emph{Commun. ACM}} \bibinfo{volume}{43},
  \bibinfo{number}{3} (\bibinfo{date}{mar} \bibinfo{year}{2000}),
  \bibinfo{pages}{45–53}.
\newblock
\showISSN{0001-0782}
\urldef\tempurl%
\url{https://doi.org/10.1145/330534.330538}
\showDOI{\tempurl}


\bibitem[Patton(1990)]%
        {patton1990qualitative}
\bibfield{author}{\bibinfo{person}{Michael~Quinn Patton}.}
  \bibinfo{year}{1990}\natexlab{}.
\newblock \bibinfo{booktitle}{\emph{Qualitative evaluation and research
  methods}}.
\newblock \bibinfo{publisher}{SAGE Publications, inc}.
\newblock


\bibitem[Quanz et~al\mbox{.}(2020)]%
        {quanz2020machine}
\bibfield{author}{\bibinfo{person}{Brian Quanz}, \bibinfo{person}{Wei Sun},
  \bibinfo{person}{Ajay Deshpande}, \bibinfo{person}{Dhruv Shah}, {and}
  \bibinfo{person}{Jae-eun Park}.} \bibinfo{year}{2020}\natexlab{}.
\newblock \showarticletitle{Machine learning based co-creative design
  framework}.
\newblock \bibinfo{journal}{\emph{arXiv preprint arXiv:2001.08791}}
  (\bibinfo{year}{2020}).
\newblock


\bibitem[Ramesh et~al\mbox{.}(2022)]%
        {ramesh2022DALLE2}
\bibfield{author}{\bibinfo{person}{Aditya Ramesh}, \bibinfo{person}{Prafulla
  Dhariwal}, \bibinfo{person}{Alex Nichol}, \bibinfo{person}{Casey Chu}, {and}
  \bibinfo{person}{Mark Chen}.} \bibinfo{year}{2022}\natexlab{}.
\newblock \bibinfo{title}{Hierarchical Text-Conditional Image Generation with
  CLIP Latents}.
\newblock
\newblock
\urldef\tempurl%
\url{https://doi.org/10.48550/ARXIV.2204.06125}
\showDOI{\tempurl}


\bibitem[Ramesh et~al\mbox{.}(2021)]%
        {pmlr-v139-ramesh21a}
\bibfield{author}{\bibinfo{person}{Aditya Ramesh}, \bibinfo{person}{Mikhail
  Pavlov}, \bibinfo{person}{Gabriel Goh}, \bibinfo{person}{Scott Gray},
  \bibinfo{person}{Chelsea Voss}, \bibinfo{person}{Alec Radford},
  \bibinfo{person}{Mark Chen}, {and} \bibinfo{person}{Ilya Sutskever}.}
  \bibinfo{year}{2021}\natexlab{}.
\newblock \showarticletitle{Zero-Shot Text-to-Image Generation}. In
  \bibinfo{booktitle}{\emph{Proceedings of the 38th International Conference on
  Machine Learning}} \emph{(\bibinfo{series}{Proceedings of Machine Learning
  Research}, Vol.~\bibinfo{volume}{139})},
  \bibfield{editor}{\bibinfo{person}{Marina Meila} {and} \bibinfo{person}{Tong
  Zhang}} (Eds.). \bibinfo{publisher}{PMLR}, \bibinfo{pages}{8821--8831}.
\newblock
\urldef\tempurl%
\url{https://proceedings.mlr.press/v139/ramesh21a.html}
\showURL{%
\tempurl}


\bibitem[Rezwana and Maher(2022)]%
        {rezwana2022AICoCreation}
\bibfield{author}{\bibinfo{person}{Jeba Rezwana} {and}
  \bibinfo{person}{Mary~Lou Maher}.} \bibinfo{year}{2022}\natexlab{}.
\newblock \showarticletitle{Understanding User Perceptions, Collaborative
  Experience and User Engagement in Different Human-AI Interaction Designs for
  Co-Creative Systems}. In \bibinfo{booktitle}{\emph{Creativity and Cognition}}
  (Venice, Italy) \emph{(\bibinfo{series}{C\&C '22})}.
  \bibinfo{publisher}{Association for Computing Machinery},
  \bibinfo{address}{New York, NY, USA}, \bibinfo{pages}{38–48}.
\newblock
\showISBNx{9781450393270}
\urldef\tempurl%
\url{https://doi.org/10.1145/3527927.3532789}
\showDOI{\tempurl}


\bibitem[Rombach et~al\mbox{.}(2022)]%
        {Rombach_2022_CVPR}
\bibfield{author}{\bibinfo{person}{Robin Rombach}, \bibinfo{person}{Andreas
  Blattmann}, \bibinfo{person}{Dominik Lorenz}, \bibinfo{person}{Patrick
  Esser}, {and} \bibinfo{person}{Bj\"orn Ommer}.}
  \bibinfo{year}{2022}\natexlab{}.
\newblock \showarticletitle{High-Resolution Image Synthesis With Latent
  Diffusion Models}. In \bibinfo{booktitle}{\emph{Proceedings of the IEEE/CVF
  Conference on Computer Vision and Pattern Recognition (CVPR)}}.
  \bibinfo{pages}{10684--10695}.
\newblock


\bibitem[Roque et~al\mbox{.}(2016)]%
        {roque2016supporting}
\bibfield{author}{\bibinfo{person}{Ricarose Roque}, \bibinfo{person}{Natalie
  Rusk}, {and} \bibinfo{person}{Mitchel Resnick}.}
  \bibinfo{year}{2016}\natexlab{}.
\newblock \showarticletitle{Supporting diverse and creative collaboration in
  the Scratch online community}.
\newblock In \bibinfo{booktitle}{\emph{Mass collaboration and education}}.
  \bibinfo{publisher}{Springer}, \bibinfo{pages}{241--256}.
\newblock


\bibitem[Saharia et~al\mbox{.}(2022)]%
        {saharia2022Imagen}
\bibfield{author}{\bibinfo{person}{Chitwan Saharia}, \bibinfo{person}{William
  Chan}, \bibinfo{person}{Saurabh Saxena}, \bibinfo{person}{Lala Li},
  \bibinfo{person}{Jay Whang}, \bibinfo{person}{Emily Denton},
  \bibinfo{person}{Seyed Kamyar~Seyed Ghasemipour},
  \bibinfo{person}{Burcu~Karagol Ayan}, \bibinfo{person}{S.~Sara Mahdavi},
  \bibinfo{person}{Rapha~Gontijo Lopes}, \bibinfo{person}{Tim Salimans},
  \bibinfo{person}{Jonathan Ho}, \bibinfo{person}{David~J Fleet}, {and}
  \bibinfo{person}{Mohammad Norouzi}.} \bibinfo{year}{2022}\natexlab{}.
\newblock \bibinfo{title}{Photorealistic Text-to-Image Diffusion Models with
  Deep Language Understanding}.
\newblock
\newblock
\urldef\tempurl%
\url{https://doi.org/10.48550/ARXIV.2205.11487}
\showDOI{\tempurl}


\bibitem[Sbai et~al\mbox{.}(2018)]%
        {Sbai_2018_ECCV_Workshops}
\bibfield{author}{\bibinfo{person}{Othman Sbai}, \bibinfo{person}{Mohamed
  Elhoseiny}, \bibinfo{person}{Antoine Bordes}, \bibinfo{person}{Yann LeCun},
  {and} \bibinfo{person}{Camille Couprie}.} \bibinfo{year}{2018}\natexlab{}.
\newblock \showarticletitle{DesIGN: Design Inspiration from Generative
  Networks}. In \bibinfo{booktitle}{\emph{Proceedings of the European
  Conference on Computer Vision (ECCV) Workshops}}.
\newblock


\bibitem[Sch{\"o}n(1984)]%
        {schon1984architectural}
\bibfield{author}{\bibinfo{person}{Donald~A Sch{\"o}n}.}
  \bibinfo{year}{1984}\natexlab{}.
\newblock \showarticletitle{The architectural studio as an exemplar of
  education for reflection-in-action}.
\newblock \bibinfo{journal}{\emph{Journal of Architectural Education}}
  \bibinfo{volume}{38}, \bibinfo{number}{1} (\bibinfo{year}{1984}),
  \bibinfo{pages}{2--9}.
\newblock


\bibitem[Sch{\"o}n(1987)]%
        {schon1987educating}
\bibfield{author}{\bibinfo{person}{Donald~A Sch{\"o}n}.}
  \bibinfo{year}{1987}\natexlab{}.
\newblock \bibinfo{booktitle}{\emph{Educating the reflective practitioner:
  Toward a new design for teaching and learning in the professions.}}
\newblock \bibinfo{publisher}{Jossey-Bass}.
\newblock


\bibitem[Seeber et~al\mbox{.}(2020)]%
        {seeber2020AITeammates}
\bibfield{author}{\bibinfo{person}{Isabella Seeber}, \bibinfo{person}{Eva
  Bittner}, \bibinfo{person}{Robert~O. Briggs}, \bibinfo{person}{Triparna {de
  Vreede}}, \bibinfo{person}{Gert-Jan {de Vreede}}, \bibinfo{person}{Aaron
  Elkins}, \bibinfo{person}{Ronald Maier}, \bibinfo{person}{Alexander~B. Merz},
  \bibinfo{person}{Sarah Oeste-Reiß}, \bibinfo{person}{Nils Randrup},
  \bibinfo{person}{Gerhard Schwabe}, {and} \bibinfo{person}{Matthias
  Söllner}.} \bibinfo{year}{2020}\natexlab{}.
\newblock \showarticletitle{Machines as teammates: A research agenda on AI in
  team collaboration}.
\newblock \bibinfo{journal}{\emph{Information \& Management}}
  \bibinfo{volume}{57}, \bibinfo{number}{2} (\bibinfo{year}{2020}),
  \bibinfo{pages}{103174}.
\newblock
\showISSN{0378-7206}
\urldef\tempurl%
\url{https://doi.org/10.1016/j.im.2019.103174}
\showDOI{\tempurl}


\bibitem[Suh et~al\mbox{.}(2021)]%
        {suh2021GenerativeMusic}
\bibfield{author}{\bibinfo{person}{Minhyang~(Mia) Suh}, \bibinfo{person}{Emily
  Youngblom}, \bibinfo{person}{Michael Terry}, {and} \bibinfo{person}{Carrie~J
  Cai}.} \bibinfo{year}{2021}\natexlab{}.
\newblock \showarticletitle{AI as Social Glue: Uncovering the Roles of Deep
  Generative AI during Social Music Composition}. In
  \bibinfo{booktitle}{\emph{Proceedings of the 2021 CHI Conference on Human
  Factors in Computing Systems}} (Yokohama, Japan) \emph{(\bibinfo{series}{CHI
  '21})}. \bibinfo{publisher}{Association for Computing Machinery},
  \bibinfo{address}{New York, NY, USA}, Article \bibinfo{articleno}{582},
  \bibinfo{numpages}{11}~pages.
\newblock
\showISBNx{9781450380966}
\urldef\tempurl%
\url{https://doi.org/10.1145/3411764.3445219}
\showDOI{\tempurl}


\bibitem[Turkle(2005)]%
        {turkle2005second}
\bibfield{author}{\bibinfo{person}{Sherry Turkle}.}
  \bibinfo{year}{2005}\natexlab{}.
\newblock \bibinfo{booktitle}{\emph{The second self: Computers and the human
  spirit}}.
\newblock \bibinfo{publisher}{Mit Press}.
\newblock


\bibitem[Wang et~al\mbox{.}(2020)]%
        {wang2020HumanAi}
\bibfield{author}{\bibinfo{person}{Dakuo Wang}, \bibinfo{person}{Elizabeth
  Churchill}, \bibinfo{person}{Pattie Maes}, \bibinfo{person}{Xiangmin Fan},
  \bibinfo{person}{Ben Shneiderman}, \bibinfo{person}{Yuanchun Shi}, {and}
  \bibinfo{person}{Qianying Wang}.} \bibinfo{year}{2020}\natexlab{}.
\newblock \showarticletitle{From Human-Human Collaboration to Human-AI
  Collaboration: Designing AI Systems That Can Work Together with People}. In
  \bibinfo{booktitle}{\emph{Extended Abstracts of the 2020 CHI Conference on
  Human Factors in Computing Systems}} (Honolulu, HI, USA)
  \emph{(\bibinfo{series}{CHI EA '20})}. \bibinfo{publisher}{Association for
  Computing Machinery}, \bibinfo{address}{New York, NY, USA},
  \bibinfo{pages}{1–6}.
\newblock
\showISBNx{9781450368193}
\urldef\tempurl%
\url{https://doi.org/10.1145/3334480.3381069}
\showDOI{\tempurl}


\bibitem[Wang and Nickerson(2017)]%
        {wang2017literature}
\bibfield{author}{\bibinfo{person}{Kai Wang} {and} \bibinfo{person}{Jeffrey~V
  Nickerson}.} \bibinfo{year}{2017}\natexlab{}.
\newblock \showarticletitle{A literature review on individual creativity
  support systems}.
\newblock \bibinfo{journal}{\emph{Computers in Human Behavior}}
  \bibinfo{volume}{74} (\bibinfo{year}{2017}), \bibinfo{pages}{139--151}.
\newblock


\bibitem[Wei et~al\mbox{.}(2022)]%
        {wei2022emergent}
\bibfield{author}{\bibinfo{person}{Jason Wei}, \bibinfo{person}{Yi Tay},
  \bibinfo{person}{Rishi Bommasani}, \bibinfo{person}{Colin Raffel},
  \bibinfo{person}{Barret Zoph}, \bibinfo{person}{Sebastian Borgeaud},
  \bibinfo{person}{Dani Yogatama}, \bibinfo{person}{Maarten Bosma},
  \bibinfo{person}{Denny Zhou}, \bibinfo{person}{Donald Metzler},
  {et~al\mbox{.}}} \bibinfo{year}{2022}\natexlab{}.
\newblock \showarticletitle{Emergent abilities of large language models}.
\newblock \bibinfo{journal}{\emph{arXiv preprint arXiv:2206.07682}}
  (\bibinfo{year}{2022}).
\newblock


\bibitem[Wu et~al\mbox{.}(2022)]%
        {wu2022AIChains}
\bibfield{author}{\bibinfo{person}{Tongshuang Wu}, \bibinfo{person}{Michael
  Terry}, {and} \bibinfo{person}{Carrie~Jun Cai}.}
  \bibinfo{year}{2022}\natexlab{}.
\newblock \showarticletitle{AI Chains: Transparent and Controllable Human-AI
  Interaction by Chaining Large Language Model Prompts}. In
  \bibinfo{booktitle}{\emph{Proceedings of the 2022 CHI Conference on Human
  Factors in Computing Systems}} (New Orleans, LA, USA)
  \emph{(\bibinfo{series}{CHI '22})}. \bibinfo{publisher}{Association for
  Computing Machinery}, \bibinfo{address}{New York, NY, USA}, Article
  \bibinfo{articleno}{385}, \bibinfo{numpages}{22}~pages.
\newblock
\showISBNx{9781450391573}
\urldef\tempurl%
\url{https://doi.org/10.1145/3491102.3517582}
\showDOI{\tempurl}


\bibitem[Yang(2020)]%
        {yang2020profiling}
\bibfield{author}{\bibinfo{person}{Qian Yang}.}
  \bibinfo{year}{2020}\natexlab{}.
\newblock \emph{\bibinfo{title}{Profiling Artificial Intelligence as a Material
  for User Experience Design}}.
\newblock \bibinfo{thesistype}{Ph.\,D. Dissertation}. \bibinfo{school}{Carnegie
  Mellon University}.
\newblock


\bibitem[Yang et~al\mbox{.}(2020)]%
        {yang2020DesignAI}
\bibfield{author}{\bibinfo{person}{Qian Yang}, \bibinfo{person}{Aaron
  Steinfeld}, \bibinfo{person}{Carolyn Ros\'{e}}, {and} \bibinfo{person}{John
  Zimmerman}.} \bibinfo{year}{2020}\natexlab{}.
\newblock \showarticletitle{Re-Examining Whether, Why, and How Human-AI
  Interaction Is Uniquely Difficult to Design}. In
  \bibinfo{booktitle}{\emph{Proceedings of the 2020 CHI Conference on Human
  Factors in Computing Systems}} (Honolulu, HI, USA)
  \emph{(\bibinfo{series}{CHI '20})}. \bibinfo{publisher}{Association for
  Computing Machinery}, \bibinfo{address}{New York, NY, USA},
  \bibinfo{pages}{1–13}.
\newblock
\showISBNx{9781450367080}
\urldef\tempurl%
\url{https://doi.org/10.1145/3313831.3376301}
\showDOI{\tempurl}


\bibitem[Yu et~al\mbox{.}(2022)]%
        {yu2022Parti}
\bibfield{author}{\bibinfo{person}{Jiahui Yu}, \bibinfo{person}{Yuanzhong Xu},
  \bibinfo{person}{Jing~Yu Koh}, \bibinfo{person}{Thang Luong},
  \bibinfo{person}{Gunjan Baid}, \bibinfo{person}{Zirui Wang},
  \bibinfo{person}{Vijay Vasudevan}, \bibinfo{person}{Alexander Ku},
  \bibinfo{person}{Yinfei Yang}, \bibinfo{person}{Burcu~Karagol Ayan},
  \bibinfo{person}{Ben Hutchinson}, \bibinfo{person}{Wei Han},
  \bibinfo{person}{Zarana Parekh}, \bibinfo{person}{Xin Li},
  \bibinfo{person}{Han Zhang}, \bibinfo{person}{Jason Baldridge}, {and}
  \bibinfo{person}{Yonghui Wu}.} \bibinfo{year}{2022}\natexlab{}.
\newblock \bibinfo{title}{Scaling Autoregressive Models for Content-Rich
  Text-to-Image Generation}.
\newblock
\newblock
\urldef\tempurl%
\url{https://doi.org/10.48550/ARXIV.2206.10789}
\showDOI{\tempurl}


\bibitem[Zhu et~al\mbox{.}(2017)]%
        {Zhu_2017_ICCV}
\bibfield{author}{\bibinfo{person}{Jun-Yan Zhu}, \bibinfo{person}{Taesung
  Park}, \bibinfo{person}{Phillip Isola}, {and} \bibinfo{person}{Alexei~A.
  Efros}.} \bibinfo{year}{2017}\natexlab{}.
\newblock \showarticletitle{Unpaired Image-To-Image Translation Using
  Cycle-Consistent Adversarial Networks}. In
  \bibinfo{booktitle}{\emph{Proceedings of the IEEE International Conference on
  Computer Vision (ICCV)}}.
\newblock


\end{thebibliography}

\appendix
\section{Appendix}

\subsection{Task Descriptions}
\label{app:task-descroption}

\subsubsection{Design prompt for the ``Moon Landing Party'' task}\hfill{}

\textbf{Assignment} You have been hired to design a poster inviting people to a fun party celebrating the 55th anniversary of the first Moon Landing. You may create multiple versions, and then choose a final design to be sent to invitees.

\textbf{About the Moon Landing} Apollo 11 (July 16–24, 1969) was the American spaceflight that first landed humans on the Moon. Commander Neil Armstrong and lunar module pilot Buzz Aldrin landed the Apollo Lunar Module Eagle on July 20, 1969. At the event, we will have Moon themed decorations, light refreshments, and very cool Moon-themed things.

\subsubsection{Design Prompt for the ``Alice's Birthday'' task}\hfill{}

\textbf{Assignment} You have been hired to design a party invitation to Alice Liddell’s birthday party.  You may create multiple versions, and then choose a final design to be sent to all of Alice’s friends.
\textbf{Who is Alice Liddell}: Alice is a fictional character and the main protagonist of Lewis Carroll's children's novel Alice's Adventures in Wonderland (1865) and its sequel, Through the Looking-Glass (1871). A child in the mid-Victorian era, Alice unintentionally goes on an underground adventure after accidentally falling down a rabbit hole into Wonderland; in the sequel, she steps through a mirror into an alternative world.

\subsubsection{Rules and Requirements}

These rules and requirements were presented before each design session. 

Please make sure your graphic follows these rules!

\begin{itemize}
    \item You may download and use graphics, images, text etc. as you see fit. 
    \item You may not use photos of faces, persons, or groups of people.
    
    \item Do not use stock images and photos, and any computer generated images of people either. 
    
    \item You may not use [the organization]’s or another company’s logo, copyrighted images, profanity, obscenity or nudity. 

    \item Your final output will be evaluated for creativity and completeness. The most creative invitations will get a bonus!
\end{itemize}

In the \Imagen{} condition, we included this statement below these rules:

WHERE APPROPRIATE, USE ENVISAGE IN THIS TASK.

\begin{landscape}

\section{Thematic coding}

\begin{table}[]
\begin{tabular}{cll}
\multicolumn{1}{l}{\textbf{Code}}                                         & \textbf{Definition}                                                                                                                      & \textbf{Example}                                                                                                                                                                                                                                                                                                                                                      \\
\rowcolor[HTML]{EFEFEF} 
\textit{Search Strategies}                                                & \begin{tabular}[c]{@{}l@{}}How participants queried a\\ search engine for images\end{tabular}                                            & \begin{tabular}[c]{@{}l@{}}``{[}Let me see{]} if I can come up with some text for happy birthday, ... \\ Another idea is search on Google for them. You switch on Creative Commons right?... \\ we can find something grab something from.'' (P1)\end{tabular}                                                                                                        \\
\textit{Prompt Strategies}                                                & \begin{tabular}[c]{@{}l@{}}How participants prompted\\ TTI model to generate images\end{tabular}                                         & \begin{tabular}[c]{@{}l@{}}``you sort of need to spend a lot of time generating a bunch of different ideas \\ and trying to figure out how to manipulate the prompt in order to get.'' (P5)\end{tabular}                                                                                                                                                              \\
\rowcolor[HTML]{EFEFEF} 
\textit{Opinionated partner}                                              & \begin{tabular}[c]{@{}l@{}}Instances when the TTI model \\ produces unexpected results\end{tabular}                                      & \begin{tabular}[c]{@{}l@{}}``Like it's always like this image...which is a great picture of a rabbit but \\ we want just {[}{]} something that looks Like in Alice in Wonderland \\ ... and I don't know how we do that.'' (P3)\\ ''Not sure how to push this into a direction that would make the drawings more useful.'' (P9)\end{tabular}                          \\
\textit{\begin{tabular}[c]{@{}c@{}}Collaboration \\ support\end{tabular}} & \begin{tabular}[c]{@{}l@{}}Instances when TTI model\\ enabled collaboration\end{tabular}                                                 & \begin{tabular}[c]{@{}l@{}}``Maybe we can try creating a cartoon image first.... \\ What do you think we should include in this? I think maybe having something from \\ Alice in Wonderland, like the Cheshire Cat.'' (P6) \\ (seeing results) ``I got pictures of cards, maybe we should get rid of \\ Alice in Wonderland and see if that helps'' (P7)\end{tabular} \\
\rowcolor[HTML]{EFEFEF} 
\textit{Artist identity}                                                  & \begin{tabular}[c]{@{}l@{}}Instances when TTI model\\ enabled participants to develop\\ their artistic \& creative identity\end{tabular} & \begin{tabular}[c]{@{}l@{}}``It's hard to control {[}{]} but also one of the cool things that it can help you \\ is  to be creative with content, like, oh, it's like made a weird cool poster thing.'' (P4)\\ ``I guess just be dropping some of these {[}images{]} into the slides'' (P11)\end{tabular}                                                            
\end{tabular}
\caption{Thematic codes and example quotes from our participants. Participant quotes have been corrected for speech ambiguities and transcription errors.}
\label{tab:challengeCodeset}
\end{table}
\end{landscape}

\end{document}